\documentclass[iop, aas_macros,numberedappendix,twocolappendix,appendixfloats]{emulateapj}
\usepackage{graphicx}
\usepackage{epstopdf}
\usepackage[usenames,dvipsnames]{color}
\usepackage{amsmath,times}

\usepackage{subfigure}
\usepackage[export]{adjustbox}

\usepackage[bookmarks=false]{hyperref}
\hypersetup{
  colorlinks= true, 
  urlcolor  = Blue, 
  filecolor=black,
  runcolor=blue,
  linkcolor = Maroon, 
  citecolor = MidnightBlue 
} 
\newcommand{\oi}{\hbox{[O\,{\sc i}]}}
\newcommand{\oiii}{\hbox{[O\,{\sc iii}]}}
\newcommand{\nii}{\hbox{[N\,{\sc ii}]}}
\newcommand{\sii}{\hbox{[S\,{\sc ii}]}}
\newcommand{\ewhd}{\hbox{EW(H$\delta$)\,$>$\,5\AA}}
\newcommand{\nad}{\hbox{Na\,{\sc i}\,D}}
\shorttitle{SPOGS I -- The Catalog}
\shortauthors{K. Alatalo et al.}

\slugcomment{Accepted to the Astrophysical Journal Supplements April 13, 2016}

\begin{document}

\title{Shocked POststarbust Galaxy Survey I: Candidate Poststarbust Galaxies with Emission Line Ratios Consistent with Shocks}
\author{Katherine Alatalo,$^{1,2}$\altaffilmark{$\dagger$} Sabrina~L. Cales,$^{3,4}$ Jeffrey~A. Rich,$^{1,2}$ Philip~N. Appleton,$^{2,5}$ Lisa J. Kewley,$^{6}$ Mark Lacy,$^{7}$ Lauranne Lanz,$^{2}$ Anne~M. Medling,$^{6}$ \& Kristina Nyland$^{7}$
}

\affil{
$^{1}$Observatories of the Carnegie Institution of Washington, 813 Santa Barbara Street, Pasadena, CA 91101, USA\\
$^{2}$Infrared Processing and Analysis Center, California Institute of Technology, Pasadena, California 91125, USA\\
$^{3}$Yale Center for Astronomy and Astrophysics, Physics Department, Yale University, New Haven, CT 06511 USA\\
$^{4}$Department of Astronomy, Faculty of Physical and Mathematical Sciences, Universidad de Concepci\'{o}n, Casilla 160-C, Concepci\'{o}n, Chile\\
$^{5}$NASA Herschel Science Center, California Institute of Technology, Pasadena, California 91125, USA\\
$^{6}$Research School of Astronomy and Astrophysics, Australian National University, Cotter Rd., Weston ACT 2611, Australia\\
$^{7}$National Radio Astronomy Observatory, 520 Edgemont Road, Charlottesville, VA 22903, USA\\
}
\altaffiltext{$\dagger$}{Hubble fellow}
\email{email:kalatalo@carnegiescience.edu}

\begin{abstract}
There are many mechanisms by which galaxies can transform from blue, star-forming spirals to red, quiescent early-type galaxies, but our current census of them does not form a complete picture. Recent observations of nearby case studies have identified a population of galaxies that quench ``quietly.''  Traditional poststarburst searches seem to catch galaxies only after they have quenched and transformed, and thus miss any objects with additional ionization mechanisms exciting the remaining gas. The Shocked POststarburst Galaxy Survey (SPOGS) aims to identify transforming galaxies, in which the nebular lines are excited via shocks instead of through star formation processes.  Utilizing the Oh-Sarzi-Schawinski-Yi (OSSY) measurements on the Sloan Digital Sky Survey Data Release 7 catalog, we applied Balmer absorption and shock boundary criteria to identify 1,067 SPOG candidates (SPOGs*) within {\em z}\,=\,0.2. SPOGs* represent 0.2\% of the OSSY sample galaxies that exceed the continuum signal-to-noise cut (and 0.7\% of the emission line galaxy sample). SPOGs* colors suggest that they are in an earlier phase of transition than OSSY galaxies that meet an ``E+A'' selection. SPOGs* have a 13\% 1.4\,GHz detection rate from the Faint Images of the Radio Sky at Twenty centimeters survey, higher than most other subsamples, and comparable only to low-ionization nuclear emission line region hosts, suggestive of the presence of active galactic nuclei. SPOGs* also have stronger \nad\ absorption than predicted from the stellar population, suggestive of cool gas being driven out in galactic winds. It appears that SPOGs* represent an earlier phase in galaxy transformation than traditionally selected poststarburst galaxies, and that a large proportion of SPOGs* also have properties consistent with disruption of their interstellar media, a key component to galaxy transformation. It is likely that many of the known pathways to transformation undergo a SPOG phase. Studying this sample of SPOGs* further, including their morphologies, active galactic nuclei properties, and environments, has the potential for us to build a more complete picture of the initial conditions that can lead to a galaxy evolving.
\end{abstract}

\clearpage

\section{Introduction}
\label{sec:Intro}

Galaxies exhibit a prominent bimodality in color, morphology, star formation rates, and stellar mass space \citep{hubble26,baade58,holmberg58,tinsley78,kauffmann+03, blanton+moustakas09}. The color-magnitude diagram is a manifestation of this bifurcation, depicting a blue cloud and a red sequence \citep{tinsley78,strateva+01,baldry+04}. The blue cloud consists of a population of galaxies that are actively star-forming, gas-rich, and disk-dominated while the red sequence is typically populated by quiescent, gas-poor galaxies with spheroidal morphologies. The total mass of blue cloud objects has remained roughly constant over the last $\sim$\,6\,Gyr (from {\em z}\,=\,1$\rightarrow$\,0), forming stars at a rate that maintains constant star formation rate\,--\,stellar mass relation, sitting on the so-called ``star formation main sequence'' \citep{noeske+07, wuyts+11}. The red sequence on the other hand has doubled in mass \citep{bell+07,bell+12}, suggesting that a non-negligible fraction of blue cloud galaxies have quenched their star formation and evolved to the red sequence. The transformation from blue to red at {\em z}\,=\,0 also appears to be one-way \citep{young+14,appleton+14}, barring an extreme external event, such as a gas-rich merger \citep{kannappan+09,mcintosh+14}. Given the multitude of pathways to red sequence, it is vitally important to understand all initial conditions that catalyze this transformation.

Galaxies located in the sparsely populated region between the blue cloud and red sequence, the so-called ``green valley'',  were hypothesized to be taking part in a rapid transition \citep{faber+07}. A recent in-depth study by \citet{schawinski+14} of the morphologies of galaxies within the green valley found that while early-type galaxies (ETG) do indeed transition rapidly ($\approx$\,100 Myrs) through the green valley, late-type galaxies (LTGs) do not. They conclude that green colors alone are insufficient to identify transitioning galaxies since LTGs inherit their green colors from a substantial buildup of an older stellar population over the course of many Gyrs while maintaining a near-constant star formation rate.

\subsection{Star Formation Quenching Mechanisms}
Many evolutionary paths to transform a blue spiral into a red elliptical have been hypothesized.  \citet{lilly+13} describes a simple picture for the evolution of a typical galaxy, beginning with a blue galaxy on the star-forming main sequence, increasing in mass by accreting gas from the cosmic web and through mergers. Once critical mass is reached, the gas supply is cut off, quenching the star formation, and the galaxy evolves onto the red sequence. This implies that the halo (bulge) grew sufficiently large to stabilize against gravitational collapse, inhibiting star formation \citep{kauffmann+03,cattaneo+06,martig+09,martig+13,davis+14}.

\citet{peng+10} suggest that in addition to morphology, environment also contributes to regulating star formation. As a galaxy falls into a cluster environment, it experiences ram pressure stripping, in which the intracluster medium strips the less dense galactic interstellar medium \citep{gunn+gott72,chung+09b}. The hot intracluster medium also plays a role in inhibiting the accretion of new neutral material \citep[strangulation;][]{bekki+02, davis+11}, inhibiting new star forming fuel from reaching the galaxy and forming stars. In the process known as harassment (high-speed fly-bys in clusters), perturbations in a galaxy's gravitational potential can affect the efficiency of star formation \citep{mihos95,moore+96,bekki98}.  

Recently, a comprehensive study of five clusters between 0.31\,$<$\,{\em z}\,$<$0.54 by \citet{dressler+13} calls into question whether the majority of galaxy transformation takes place after the galaxies have been accreted by, and virialized with the cluster, instead suggesting that group pre-processing (during the fall-in group phase) plays a larger role in the late-type to early-type transitions than the cluster itself. In fact, tidal stripping and harassment are also at play in compact groups \citep{hickson97,verdes-montenegro+01,rasmussen+08,sivanandam+10}, and observational evidence suggests that compact group galaxies evolve rapidly \citep{johnson+07,walker+10}, thus the possibility that cluster galaxies could be transformed in the fall-in group phase rather than virialized within the cluster is plausible.

In the field, mergers and interactions play a prominent role in galaxy evolution \citep{toomre72,springel+05,hopkins+06}, transforming disks to spheroids, driving massive amounts of gas to nuclear regions, and triggering star formation and black hole growth. This activity rapidly depletes available gas stores via the starburst and can trigger  feedback from the active galactic nucleus \citep[AGN;][]{dimatteo+05,feruglio+10,cicone+14,cicone+15}, quenching star formation. Observational evidence supports the merger hypothesis: the most luminous AGN tend to be produced by mergers while less luminous AGN are mainly powered by secular processes \citep{treister+12}.  AGN and star formation can also provide negative feedback that regulates star formation. Radiative-mode AGN can cause ionization, heating, and radiation pressure that may regulate star formation \citep{ciotti+ostriker07,cano-diaz+12}. Mechanical-mode AGN that drive nuclear winds and jets can remove fuel for star formation from the galaxy \citep{birzan+08,ciotti+10,alatalo+11,nyland+13}, though recent simulations and observations seem to indicate AGN feedback mechanisms at {\em z}\,=\,0 do not provide the global star formation quenching that were expected from previous simulations \citep{springel+05}, but rather act quite locally ($\lesssim$\,1\,kpc) to the supermassive black hole \citep{garcia-burillo+14,costa+14,gabor+14,alatalo15}. However, there is evidence that quasars at high-{\em z} produce the extreme energies required to change the global properties of their hosts \citep{zakamska+15}.

\subsection{Quiet Quenching}
The nearby, low ionization nuclear emission line region (LINER; \citealt{kewley+06}) galaxy, NGC 1266, although unremarkable in its ground-based optical appearance, was found to be hosting a dramatic AGN-driven multiphase outflow, and a 10$^9$\,M$_\odot$ reservoir of molecular gas contained in the nucleus \citep{alatalo+11}.  Upon deeper inspection, NGC\,1266 was found to contain shocked ionized gas ratios associated with the outflow \citep{davis+12}, a Compton-thick AGN \citep{nyland+13,a15_sfsupp}, and an intermediate stellar population (1/2\,Gyr; \citealt{a14_stelpop}). The star formation within the molecular core of the galaxy was also found to be severely suppressed, leading to an inefficiency of $\gtrsim$\,50 \citep{a15_sfsupp}, likely due to the turbulence being injected into the system by the powerful shocks associated with the outflow, thereby extending the lifetime of the nuclear molecular gas by two orders of magnitude. \citet{a14_stelpop} concluded that the observational properties of NGC\,1266 were likely caused by a minor merger that drove gas to the nuclear region of the galaxy and triggered a burst of star formation but that these events occurred 1/2\,Gyr ago, and have left few signs of the initial triggering event in the galaxy.  Ultimately, the molecular disk fueled the black hole, which formed a low-power radio jet that quenched the initial starburst and continues to inhibit star formation through turbulence, rather than directly expelling the gas from the halo \citep{a15_sfsupp}. 

Indeed the recent discoveries into the nature of NGC\,1266, a galaxy undergoing substantial transformation and star formation quenching, suggests that there is a population of objects that have not previously been identified as transforming, due to their lack of obvious clues such as tidal tails, starburst signatures, or prominent AGN activity. We call this particular pathway ``quiet quenching.''  NGC\,1266 is an ideal case study for this new, quietly quenching population, which transform from spirals into ellipticals in spite of there being no outward, obvious signs of the process.  Thus, galaxies with poststarburst spectral signatures and shock-like ionized gas line ratios may comprise an important, overlooked segment of the transitioning galaxy population, containing exactly those objects which are being actively impacted by turbulence.

\subsection{Shocked Poststarburst Galaxies (SPOGs)}
Many of the outlined mechanisms that lead to the transformation of a galaxy are violent, and thus are accompanied by shocks that inject turbulence into the star-forming gas, heating it and serving as a catalyst for its expulsion, ultimately quenching star formation in the form of a massive galactic wind \citep{armus+90,heckman+90,rich+10,sturm+11}.  Shocks are prevalent in merging systems \citep{rich+11,soto+12,inami+13,rich+14,peterson+12},  in galaxies located in the outskirts of clusters \citep{braglia+09,sivanandam+10,cen+14}, and galaxy groups \citep{appleton+06, cluver+10, appleton+13, cluver+13,vogt+13,vogt+15}.  Shocks are also a ubiquitous feature of galactic winds that are being driven by quasars \citep{nesvadba+08,harrison+14,villar-martin+14}.  
For transitioning galaxies in both the modern and early universe, shocks serve as a beacon pinpointing those galaxies that are in the most dramatic phases of their evolution.

One classical way that galaxies have been identified as transitioning objects is by searching for poststarburst galaxies \citep{dressler+gunn83,zabludoff+96}, which stood out based on their disparate emission line and stellar population characteristics.  Poststarburst galaxies are identified based on the {\em presence} of an intermediate age (A-star) stellar population and the {\em absence} of emission lines consistent with star formation (using nebular lines such as H$\alpha$ or [O\,{\sc ii}]$\lambda$3727\AA; \citealt{quintero+04,goto05,goto07}). This disparate set of characteristics demonstrates that these objects are a subset that had a recent burst of star formation that has ceased rapidly.  This type of search was used as the benchmark for finding transitioning objects, and yet a classical poststarburst search would miss the massive molecular outflow host NGC\,1266 due to its shock-powered H$\alpha$ emission \citep{davis+12,a14_stelpop}.

New studies expanding the selection criteria for poststarburst galaxies have called into question the reasoning behind the limit placed on nebular emission \citep{falkenberg+09,yesuf+14}. In particular, placing a limit on nebular emission lines introduces a bias against AGNs \citep{wild+09,kocevski+11,cales+11,cales+13,a14_irtz} because AGN can power significant emission of [O\,{\sc ii}] and H$\alpha$. Placing a limit on nebular emission biases the poststarburst selection against any  quenching objects that contain energetic phenomena (such as AGNs or shocks) capable of exciting those lines. Furthermore, an [O\,{\sc ii}]  limit also produces a metallicity bias, since [O\,{\sc ii}] is weak at both very high and very low metallicities \citep{kewley+04}. 

A secondary limitation of the poststarburst search is that by requiring the absence of [O\,{\sc ii}] and H$\alpha$, it is only able to identify galaxies after all star formation has ceased, rather than at the point star formation is abruptly diminishing, increasing the time between identification and the suspected preceding starburst phase \citep{hogg+06}. \citet{snyder+11} investigated the quenching process and the poststarburst phase in a suite of simulations, which showed substantial scatter in the total time of the K+A phase, but also significant dependencies on many of the initial conditions of the merger. By the time the galaxy has a detectable K+A phase, many of the initial conditions associated with the triggering event have faded. Therefore, identifying galaxies at an earlier phase of transition is also important to a complete understanding of how galaxies quench, and transform from blue to red.

\begin{table}[t!]
\centering
\caption{Line Diagnostic Type} \vspace{1mm}
\begin{tabular}{l r r}
\hline \hline
{\bf Type} & {\bf Number} & {\bf Percentage}$^\dagger$\\
\hline
Ambiguous & 17,097 & 10.7$\pm$0.07\\
SF & 111,972 & 70.2$\pm$0.11\\
Composite & 14,226 & 8.9$\pm$0.07\\
Seyfert & 4,765 & 3.0$\pm$0.04\\
LINER & 11,327 & 7.2$\pm$0.06\\
\hline
{\bf Total} & 159,387 & 100.0\\
\hline \hline
\end{tabular} \\
\label{tab:bpt}
\raggedright {\footnotesize
$^\dagger$Includes binomial errors contribute negligibly.
}
\vspace{1mm}
\end{table}
\begin{figure}[t]
\centering
\includegraphics[width=0.45\textwidth,clip,trim=1.5cm 2.7cm 1cm 1.6cm]{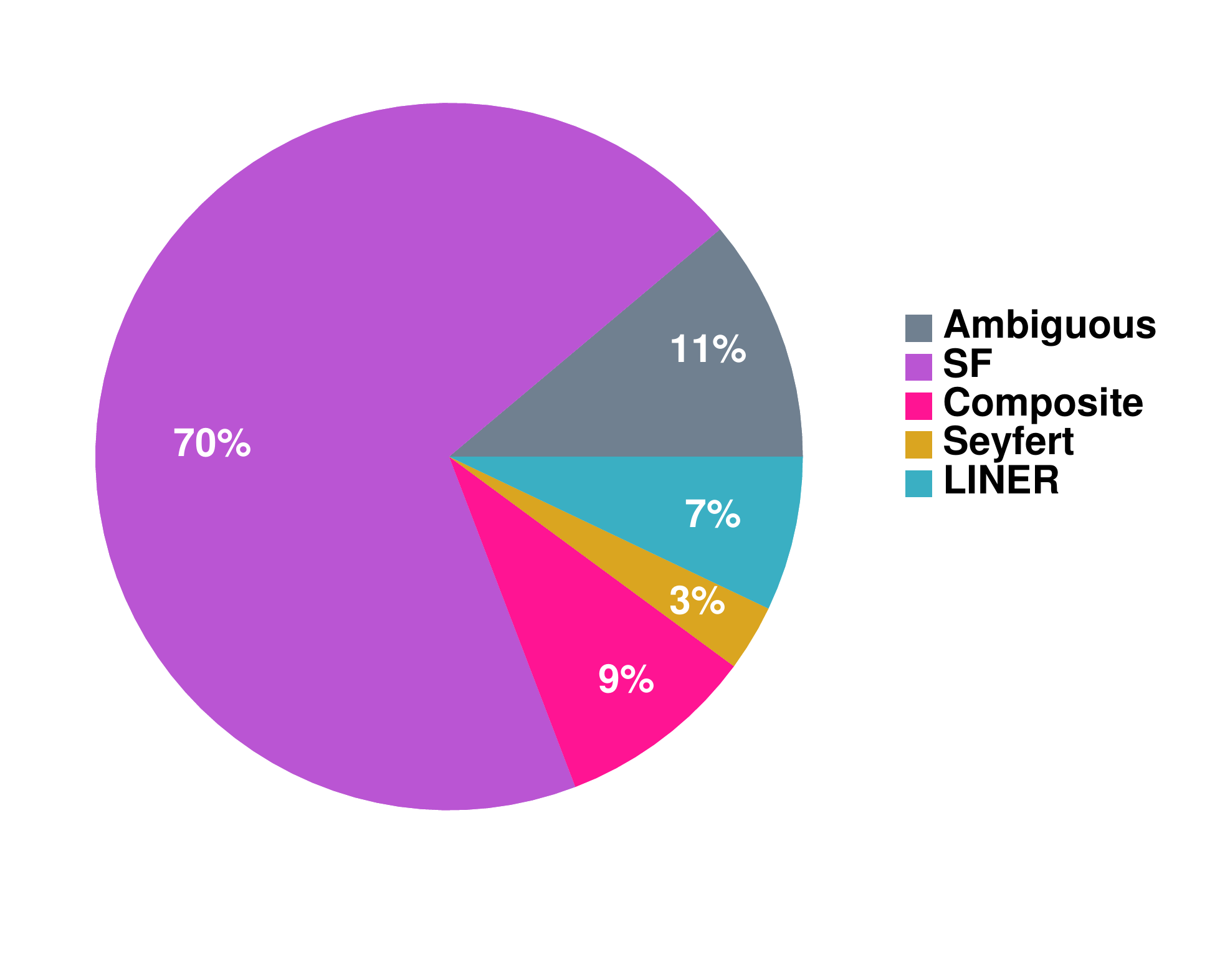}
\caption{\footnotesize The fractional representation of objects in the emission line galaxy (ELG) sample. The majority (70\%) of galaxies in the ELG contain line diagnostic ratios consistent with star formation.}
\label{fig:bpt_pie}
\end{figure}

If quenching galaxy searches reject NGC\,1266-like galaxies, they may miss a substantial population of transitioning galaxies.  Given that, at {\em z\,=}\,0, the evolutionary path from blue spiral to red early-type is one-way \citep{appleton+14, young+14}, it is essential to understand {\em all} initial conditions that could lead to this transition, including the population that has quietly quenched. Robustly measuring the fraction of quiet quenchers occurring in the overall galaxy population will provide an estimate of their duty cycle, and studying them in detail will provide insights into the connections between the source of the narrow line emission and the quenching of star formation. To this end, we have created the Shocked POststarbust Galaxy Survey (SPOGS\footnote{\href{http://www.spogs.org}{http://www.spogs.org}}), building a catalog of poststarburst galaxies hosting narrow line ratios consistent with shocks. This catalog focuses on Sloan Digital Sky Survey (SDSS) DR7 \citep{sdssdr7}\footnote{\href{http://classic.sdss.org/dr7/}{http://classic.sdss.org/dr7/}} galaxies with {\em z}\,$<$\,0.2 and utilizes the Oh-Sarzi-Schawinski-Yi absorption and emission line catalog (OSSY; \citealt{ossy})\footnote{\href{http://gem.yonsei.ac.kr/~ksoh/wordpress/}{http://gem.yonsei.ac.kr/$\sim$ksoh/wordpress/}} .

This paper is part of a series dedicated to studying the connection between the total emission-line galaxy population and the intersection of transitioning galaxies and those with shock line ratios. In \S\ref{sec:sample}, we lay out the selection criteria for the SPOG sample and present the parent sample. In \S\ref{sec:disc}, we discuss the role SPOGs play in the context of galaxy evolution and lay groundwork for deeper investigations of SPOGs. In \S\ref{sec:sum}, we summarize our results. The cosmological parameters $H_0 = 70~$km~s$^{-1}$, $\Omega_m = 0.3$ and $\Omega_\Lambda = 0.7$ \citep{wmap} are used throughout.

\section{SPOGS Sample Definition and Characteristics}
\label{sec:sample}

\subsection{The Emission Line Galaxy Catalog}
\label{sec:catalog}

To build a catalog of SPOG candidates (deemed SPOGs*), we drew from the OSSY sample, a comprehensive database of the emission, absorption, and continuum measurements of 664,187 Sloan Digital Sky Survey (SDSS) Data Release 7 \citep[DR7;][]{sdssdr7} galaxies within {\em z}\,=\,0.2.\footnote{Approximately 7\% of SDSS objects have multiple spectra in the SDSS database, so the spectra do not necessarily represent unique objects.} OSSY uses penalized pixel-fitting algorithms ({\sc ppxf}; \citealt{ppxf}) to fit stellar population templates and the absorption features and kinematics in combination with an algorithm that fits the absorption and emission lines ({\sc gandalf}; \citealt{sarzi+06}).
\begin{figure}[t]
\raggedright
\includegraphics[width=0.485\textwidth,clip,trim=0cm 0cm 0cm 0cm]{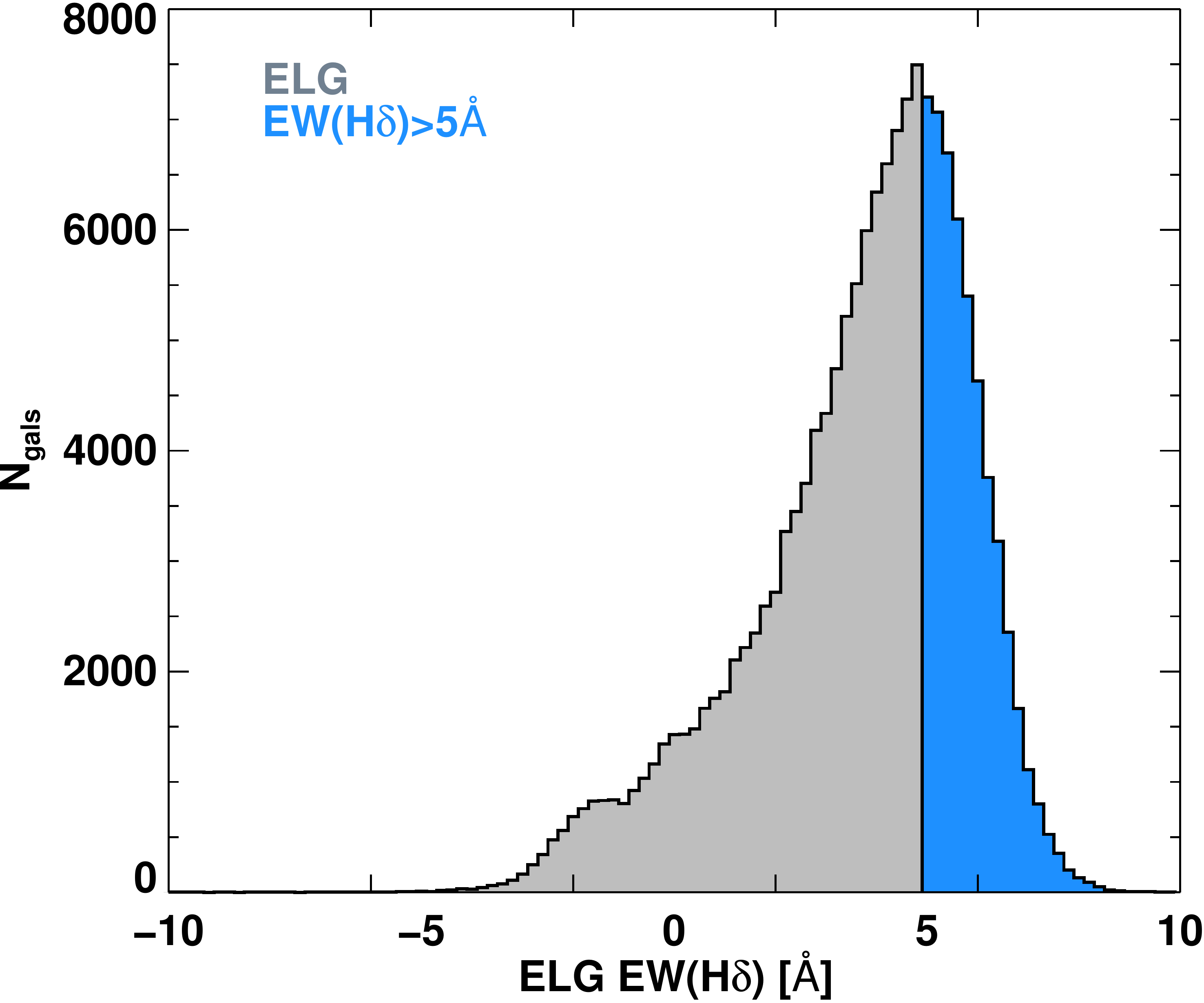}
\caption{The distribution of EW(H$\delta$) values in the ELG sample. 29\% of galaxies in the ELG sample have \ewhd, shaded in blue.}
\label{fig:hdelta}
\end{figure}

Here we describe some of the fit and quality assurance parameters from the OSSY measurements \citep{ossy}. The signal, statistical noise and residual noise are measured using the averages of the continuum bands 4500-4700\,\AA, 5400-5600\,\AA, 6000-6200\,\AA, and 6800-7000\,\AA. The ratio of the fit residuals (i.e., residual noise, {\em rN}) to statistical fluctuations ({\em sN}) corresponds to the reduced $\chi^2$ statistical measurement and is used to assess the quality of the model. When binned by the signal-to-noise (S/N) this goodness-of-fit parameter ({\em rN/sN}) centers around unity with a distribution of values above and below this value of standard deviation, $\sigma$. Objects far from unity, as measured by the number of $\sigma$ from the median {\em rN/sN} ($N_{\sigma}$) have particularly poor fits.

The parent Emission Line Galaxy (ELG) catalog is a subset of the OSSY catalog that contains all emission line galaxies that satisfy our continuum and emission-line S/N criteria.  To build the parent sample, we limited the ELG catalog to data from the OSSY database meeting the following 2 quality assurance criteria in continuum:

\begin{itemize}
\item S/N\,$>$\,10, to ensure robust absorption feature detection
\item {\em N}$_\sigma$\,$<$\,3, which limits the ratio of residual to statistical noise in units of standard deviations from the median
\end{itemize}

591,627 (89\%) galaxies meet the S/N continuum criteria. We additionally applied the following quality assurance criteria for {\em all} the following narrow lines: H$\beta$, \oiii, H$\alpha$, \nii, \sii, and \oi:

\begin{itemize}
\item A/N\,$>$\,1, peak line flux (amplitude) to noise
\item Line {\em N}$_\sigma$\,$<$\,5, which limits the ratio of residual to statistical noise in units of standard deviations from the median
\end{itemize}

These criteria ensure that our parent sample includes only robust detections of both absorption and emission lines, and allows us to classify galaxies over the entire suite of line diagnostics (\oiii/H$\beta$ versus \nii/H$\alpha$, \sii/H$\alpha$, and \oi/H$\alpha$), reducing ambiguity.  A total of 159,387 galaxies (24$\pm$0.05\%) of the OSSY sample pass all of our quality control criteria and are considered part of the parent sample.  The ELG sample was then sub-classified based on their line diagnostic ratios \citep{bpt,veilleux+87} as either Seyferts, LINERs, composites, star-forming or ambiguous (in which an object did not fit into a single classification across all three line diagnostics) based on the models and empirical classification lines of \citet{kewley+06}.  Table\,\ref{tab:bpt} and Figure\,\ref{fig:bpt_pie} show the fractional representation of each line diagnostic state found in the ELG sample.  It is clear that the majority of objects that passed quality control (and thus had strong line emission) are star-forming galaxies.

\subsection{Stellar Population Criterion}

Classically a poststarburst galaxy is defined by \citet{dressler+gunn83} as one that exhibits strong Balmer absorption lines, indicating intense star formation in the past $\sim$\,Gyr, and a lack of ongoing star formation, as indicated by having little or no nebular emission (usually chosen to be either the H$\alpha$ or [O\,{\sc ii}]$\lambda$3727\AA\ line).  Their spectra are best modeled as a superposition of an elliptical galaxy-like spectrum and an A-star spectrum, hence they are often called ``E+A'' or ``K+A'' galaxies \citep{quintero+04}.  Balmer absorption is strongest in A-type stars \citep{vazdekis+10}, and thus is often used as a tracer of a stellar population that was formed in a recent burst of star formation. \citet{falkenberg+09} calculated a grid of galaxy models with different combinations and a sudden increase or decrease of the star formation rate on different time-scales and find that a poststarburst phase can be induced via starbursts, or an abrupt termination of star formation. We have chosen the Balmer line, H$\delta$, for our cut, since it is neither contaminated by absorption  lines (such as H$\epsilon$, which falls within the Ca\,{\sc ii}\,H absorption trough), nor substantially filled in by nebular emission lines (as is common in H$\alpha$ and non-negligible in H$\beta$ and H$\gamma$). The OSSY catalog uses emission line infilling-subtracted absorption line fits to mitigate the effects of infilling, though there likely still persist biases against galaxies with strong emission lines, such as AGNs, and towards more highly reddened sources. We set a threshold of \ewhd, consistent with the Balmer poststarburst criteria of \citet{goto07} and \citet{falkenberg+09}.

\begin{figure}
\centering \vskip -1mm
\subfigure{\includegraphics[width=0.47\textwidth]{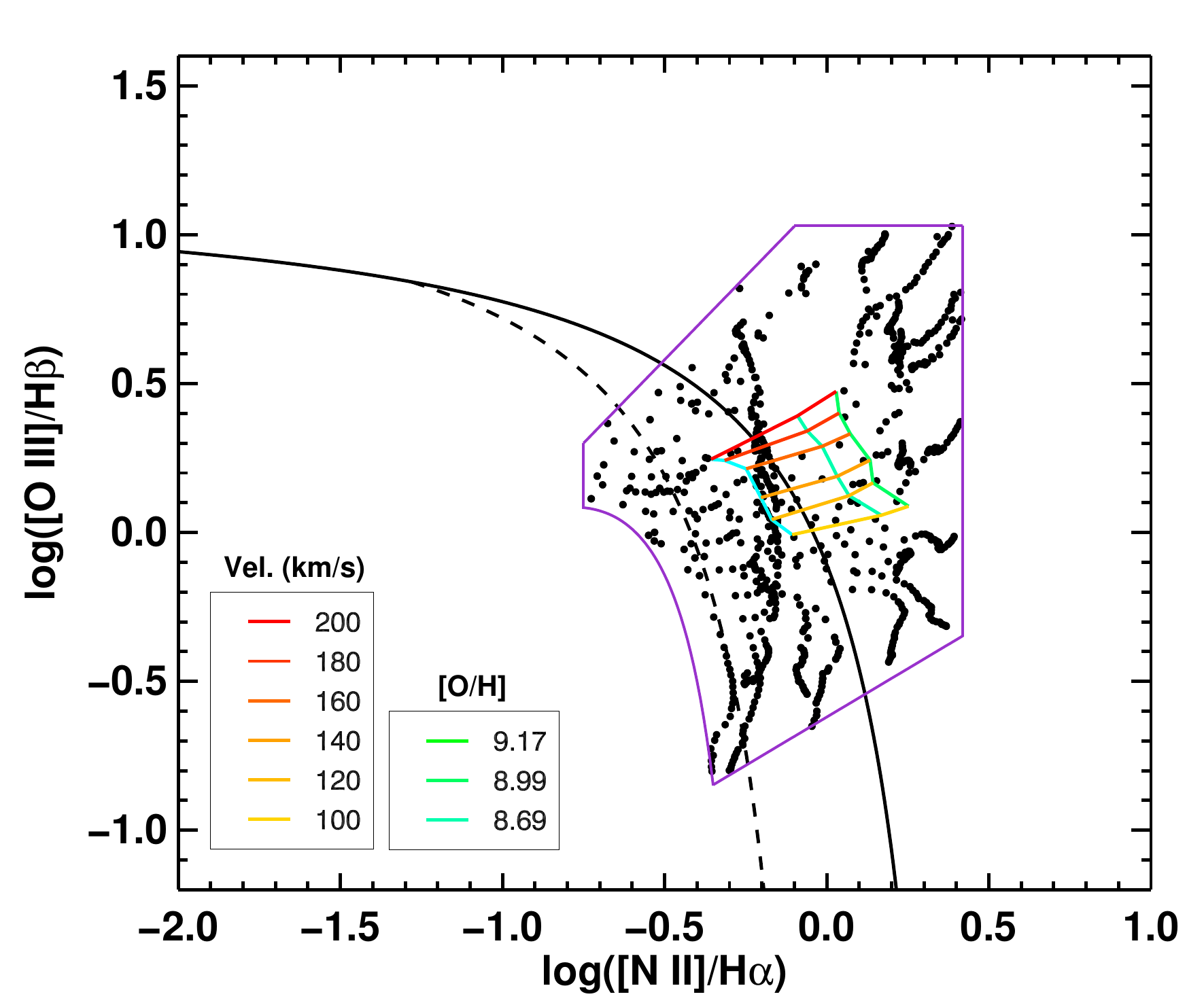}}\vskip -5mm
\subfigure{\includegraphics[width=0.47\textwidth]{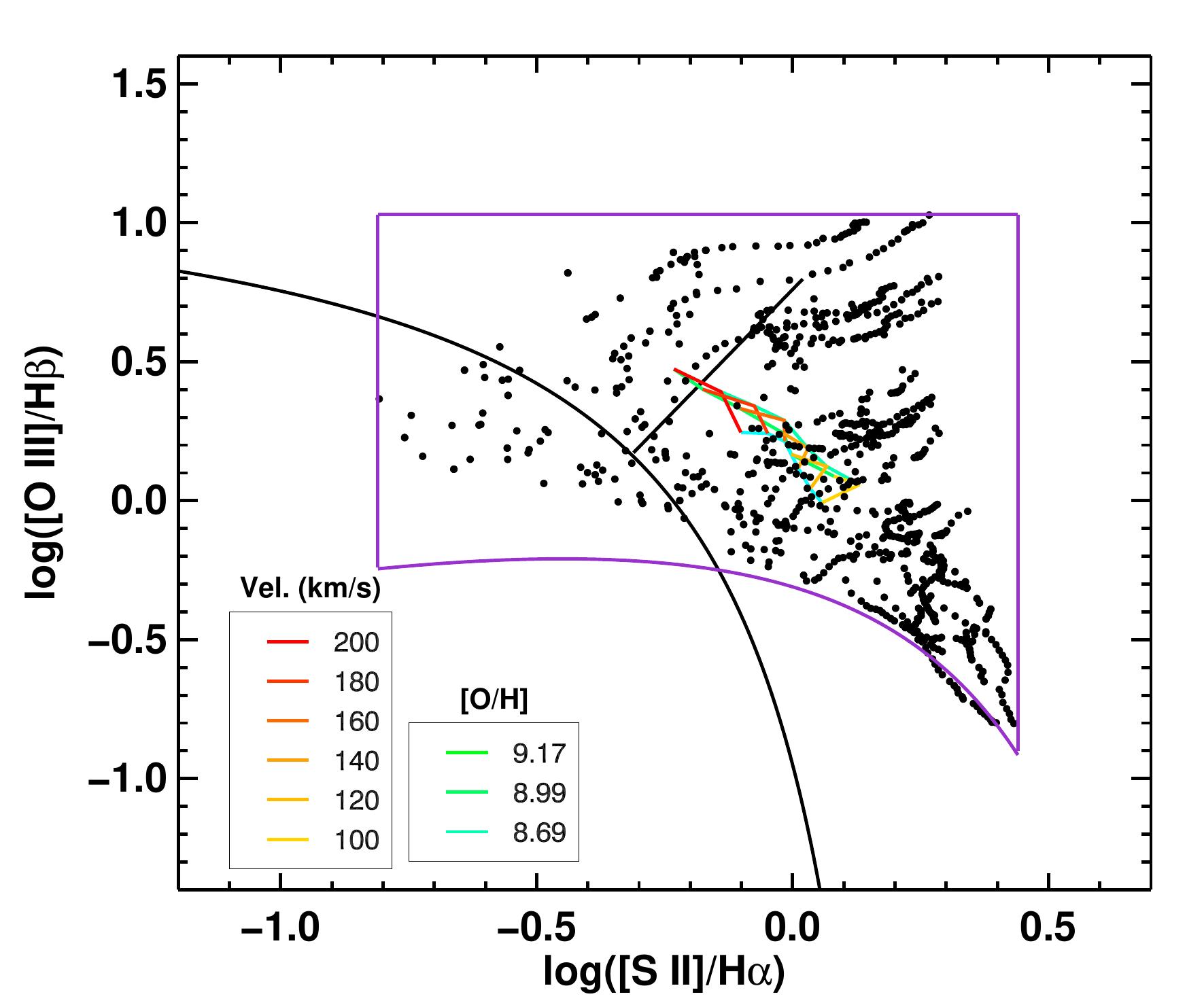}} \vskip -5mm
\subfigure{\includegraphics[width=0.47\textwidth]{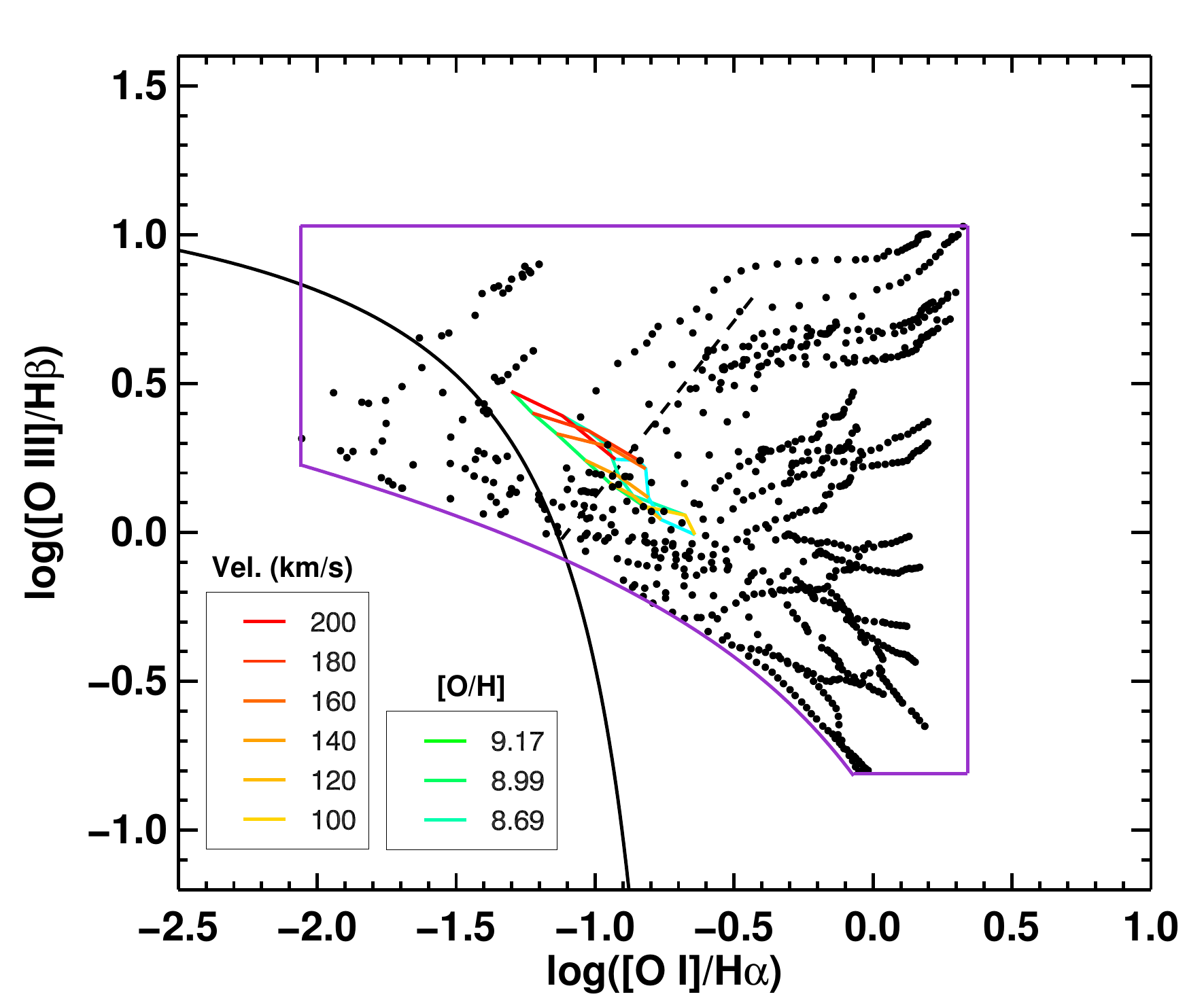}} \vskip -3mm
\caption{\footnotesize Line diagnostic diagrams  \citep{bpt,veilleux+87} for \oiii/H$\beta$ versus \nii/H$\alpha$ (top), \sii/H$\alpha$ (middle), and \oi/H$\alpha$ (bottom) with both fast (black points; \citealt{allen+08}) and slow (colored grids; \citealt{rich+11}) shock models with various metallicities, and shock velocities ranging from 100--1000 km s$^{-1}$, overlaid with the star formation boundary (black line; \citealt{kewley+06}, dashed line in \nii/H$\alpha$; \citealt{kauffmann+03}).  The fast shock grids cover a much broader range of velocities, resulting in a correspondingly larger range of permissible emission line ratios. The purple line defines the boundaries put in place to encompass the entire phase space of the shock models (defined in \S\ref{sec:shocks}).  These models show that shocks can be located in all regions of the diagnostic diagrams, mimicking line ratios associated with SF, composites, Seyferts, and LINERs.}
\label{fig:shockmodels}
\end{figure}

Figure\,\ref{fig:hdelta} shows the 46,936 (29$\pm$0.11\%) objects in the ELG that have \ewhd. 
The majority of sources in the ELG sample have line diagnostics consistent with star formation (Fig.\,\ref{fig:bpt_pie}). \citet{a14_irtz} also showed that most sources with \ewhd\ had very blue colors.  In fact, 44,327 (94$\pm$0.10\%) of galaxies that fit the \ewhd\ criteria also lie in the star-forming region of the line diagnostic diagrams.  This is indicative of a population that is continuously forming stars, having built a substantial intermediate stellar population along the way. 

\subsection{Colors and Stellar Masses}
\label{sec:colors}
We use the dereddened magnitudes from SDSS DR9 \citep{sdssdr9} (which are corrected for Galactic extinction) for our bandpasses. The {\em u--r} colors are then {\em k}-corrected using the {\sc idl} routine {\tt calc\_kcor}\footnote{\href{http://kcor.sai.msu.ru/}{http://kcor.sai.msu.ru/}} \citep{calc_kcor}, and corrected for intrinsic extinction using the stellar E(B--V) values from the OSSY catalog \citep{ossy}. The stellar masses were determined from the {\em k}-corrected, extinction corrected {\em i}-band dereddened fluxes and {\em u--r} colors, using the $M_i$\,-\,{\em u--r} to stellar mass conversion from \citet{bell+03}.

\subsection{Defining the Shock Boundaries}
\label{sec:shocks}

To search for shocked gas in galaxies, we rely on grids of shock models generated with {\sc mappings\,iii} \citep{dopita+sutherland93,dopita+96,dopita+05,allen+08,rich+11}.  Figure\,\ref{fig:shockmodels} show the distribution of fast (black points; \citealt{allen+08}) and slow (colored lines; \citealt{rich+11}) shock model grids in the line diagnostic diagrams with log(\oiii/H$\beta$) versus log(\nii/H$\alpha$), log([\sii/H$\alpha$), and log(\oi/H$\alpha$) optical emission line ratios. The fast shock grids cover a much broader range of velocities (from 100--1000\,km~s$^{-1}$), with the fast shock models \citep{dopita+05,allen+08} covering 1\,{\em Z}$_\odot$ and 2\,{\em Z}$_\odot$ metallicities, and slow shock models covering 1,\,2, and 3\,{\em Z}$_\odot$ metallicities. \oi/H$\alpha$ is a particularly good tracer of shock excitation \citep{farage+10,rich+10}. The possible line ratios cover a wide excitation space, and we have used these distributions to define a polygon that encompass galaxy emission-line ratios consistent with shock excitation (Figure\,\ref{fig:shockmodels}). The area covered by these polygons also overlaps regions traditionally associated with other emission-line processes including LINER, Seyfert, and H\,{\sc ii} region emission \citep{kauffmann+03,kewley+06}. This is unavoidable, but is not unexpected given observations of shocks in nearby galaxies (e.g. \citealt{rich+11,ho+14}).  The shock criteria and boundaries are defined based on the following equations:\\

\noindent Shock criteria:
\begin{equation}
-0.75\ <\ \log([\rm{N\ II}]/\rm{H\alpha})\ <\ 0.42
\end{equation}
\begin{equation}
-0.81\ <\ \log([\rm{S\ II}]/\rm{H\alpha})\ <\ 0.44
\end{equation}
\begin{equation}
-2.06\ <\ \log([\rm{O\ I}]/\rm{H\alpha})\ <\ 0.34
\end{equation}
\begin{equation}
-0.81\ <\ \log([\rm{O\ III}]/\rm{H\beta})\ <\ 1.03
\end{equation}

and functions:

\begin{equation}
\begin{split}
\log([\rm{O\ III}]/\rm{H\beta})\ >\ 0.4/[\log([\rm{N\ II}]/\rm{H\alpha})\ +\ 0.15]\ +\ \\
\log([\rm{N\ II}]/\rm{H\alpha})\ +\ 1.5
\end{split}
\end{equation}
\begin{equation}
\log([\rm{O\ III}]/\rm{H\beta})\ >\ 0.65\log([\rm{N\ II}]/\rm{H\alpha})\ -\ 0.62
\end{equation}
\begin{equation}
\log([\rm{O\ III}]/\rm{H\beta})\ <\ 1.12\log([\rm{N\ II}]/\rm{H\alpha})\ +\ 1.14
\end{equation}
\begin{equation}
\begin{split}
\log([\rm{O\ III}]/\rm{H\beta})\ >\ 1.05/[\log([\rm{S\ II}]/\rm{H\alpha})\ -\ 1.00]\ +\ \\
0.5\log([\rm{S\ II}]/\rm{H\alpha})\ +\ 0.74
\end{split}
\end{equation}
\begin{equation}
\begin{split}
\log([\rm{O\ III}]/\rm{H\beta})\ >\ 1.15/[\log([\rm{O\ I}]/\rm{H\alpha})\ -\ 0.95]\ -\ \\
0.15\log([\rm{O\ I}]/\rm{H\alpha})\ +\ 0.30
\end{split}
\end{equation}

\begin{figure*}[t]
\centering
\subfigure{\includegraphics[width=0.33\textwidth]{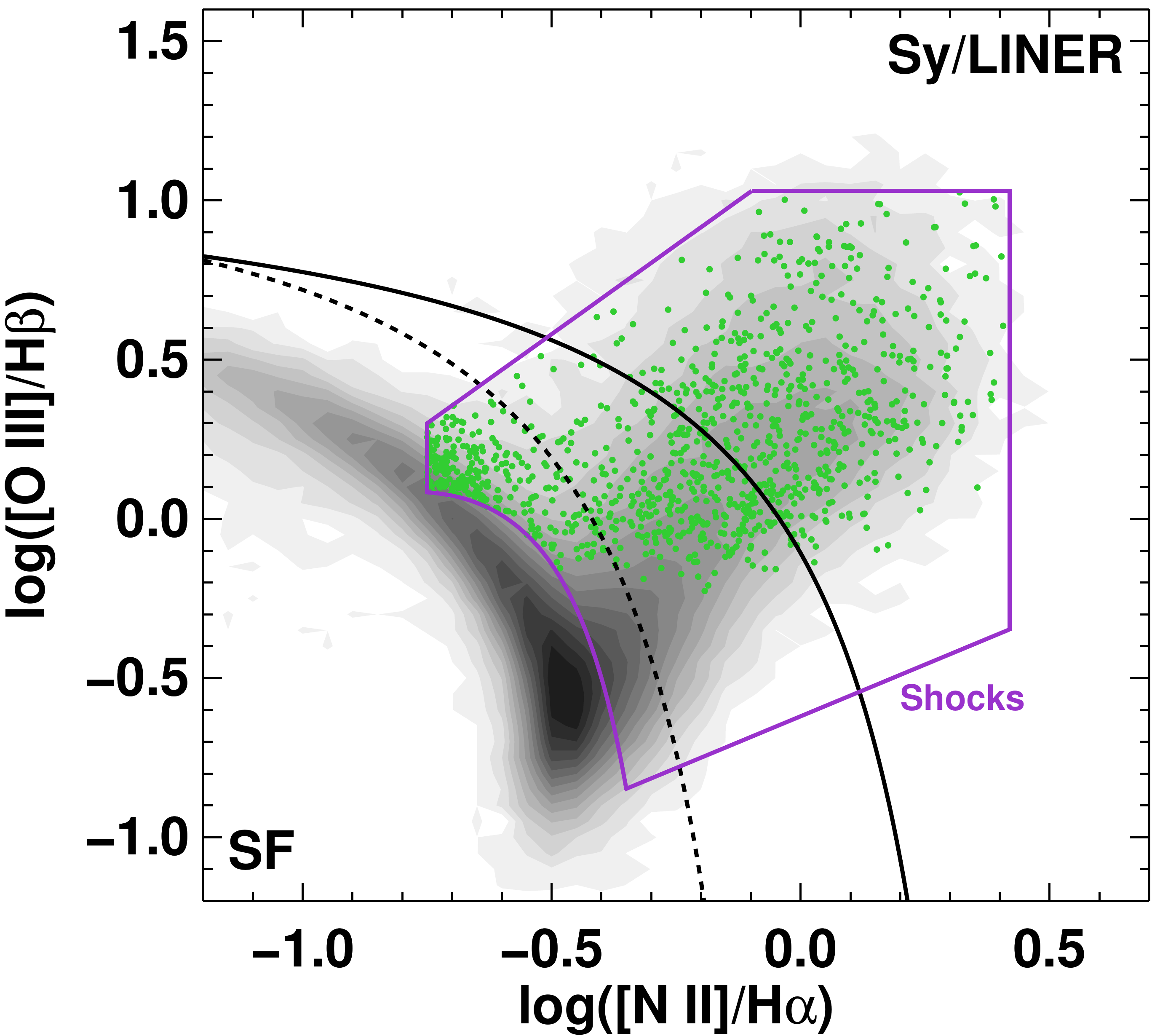}}
\subfigure{\includegraphics[width=0.33\textwidth]{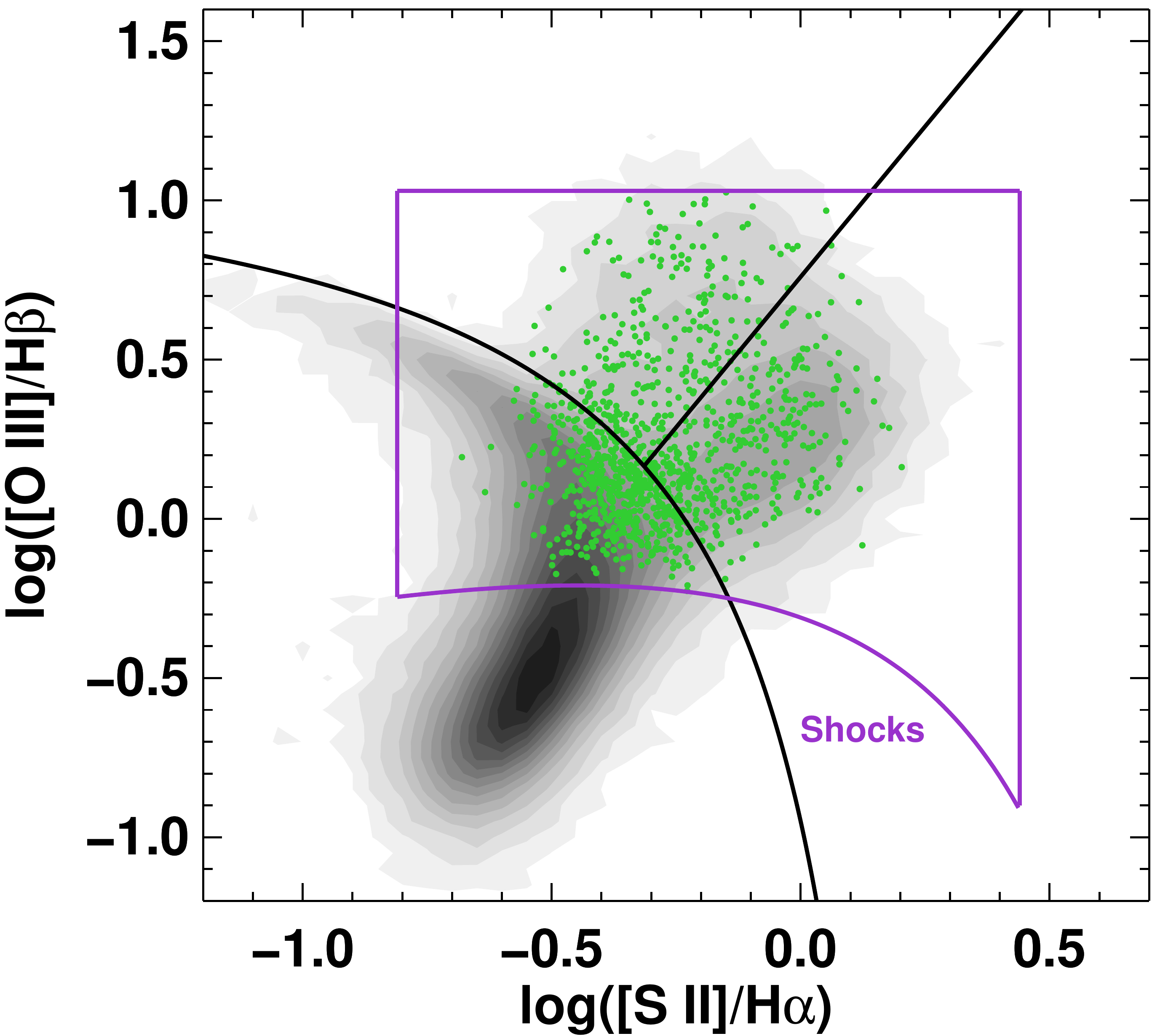}}
\subfigure{\includegraphics[width=0.33\textwidth]{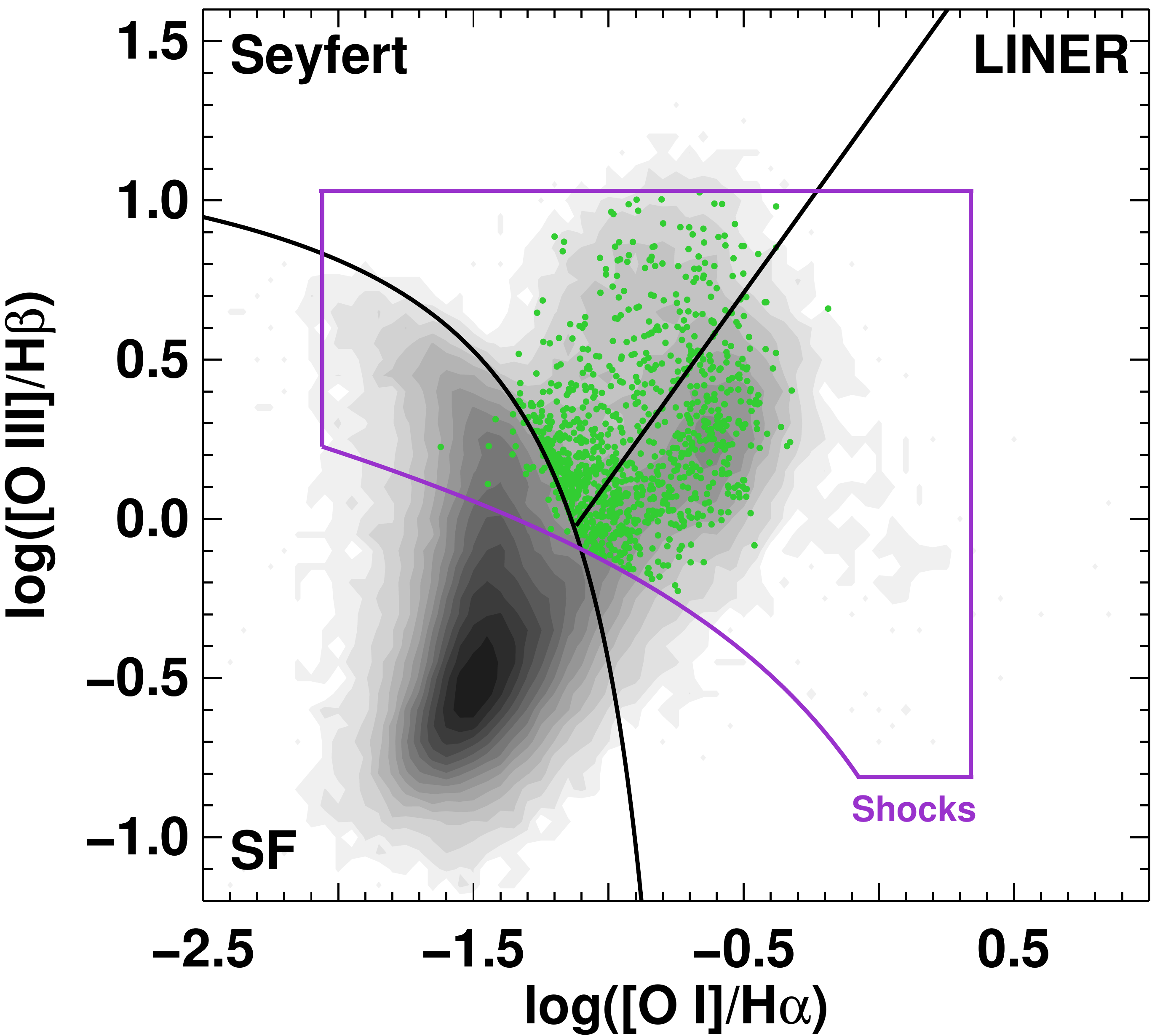}}
\caption{\footnotesize The line diagnostic diagrams \citep{bpt,veilleux+87} for \oiii/H$\beta$ versus \nii/H$\alpha$ (left), \sii/H$\alpha$ (center), and \oi/H$\alpha$ (right) are overlaid with the ELG sample (grayscale) and SPOGs* (green points).  Labels on \nii/H$\alpha$ and \oi/H$\alpha$ label the locations of different ionizing mechanisms from \citet{kewley+06} demarcated by black lines (the dashed line in the \nii/H$\alpha$ plot is from \citealt{kauffmann+03}). The composite region is shown between the dashed and solid black line on the \nii/H$\alpha$ diagram.  Shock boundaries are overlaid in purple. SPOGs* populate a large fraction of the theoretically possible locations of shocks.}
\label{fig:spog_bpt}
\end{figure*}

\begin{table}[h!]
\centering
\caption{Candidate SPOG Catalog} \vskip -3mm
\rotatebox{90}{
\begin{tabular}{l c r r r r r r r r r r r r r r r r}
\hline \hline
{\bf SPOG} & {\bf IAU Name} & {\bf Redshift} & \textbf{\em u} & \textbf{\em g} & \textbf{\em r} & \textbf{\em i} &\textbf{\em z} &
\textbf{\em M$_r$} & \textbf{log(M$_\star$)} & \textbf{H$\alpha$} & \textbf{H$\beta$} & \textbf{\oiii} & \textbf{\nii} & \textbf{\sii} & \textbf{\oi} &
{\bf Object} & {\bf Spectrum}\\
{\bf ID} & {\bf (SDSS)} & & & & & & & & {\bf (M$_\odot$)}& & & & & & & {\bf Link} & {\bf Link} \\
{\bf (1)} & {\bf (2)} & {\bf (3)} & {\bf (4)} &  {\bf (5)} & {\bf (6)} & {\bf (7)} & {\bf (8)} & {\bf (9)} & {\bf (10)} & {\bf (11)} & {\bf (12)} & {\bf (13)} & {\bf (14)} & 
{\bf (15)} & {\bf (16)} & {\bf (17)} & {\bf (18)} \\
\hline
       1 & J000318.22+004844.3 & 0.1389 & 19.75 & 18.26 & 17.46 & 17.04 & 16.72 & -21.87 & 10.62 & 2.29 & 1.32 & 1.90 & 2.23 & 1.78 & 1.32 & \href{http://skyserver.sdss.org/dr9/en/tools/explore/obj.asp?ra=0.8259&dec=0.8123}{Object} & \href{http://dr9.sdss3.org/spectrumDetail?plateid=387&mjd=51791&fiber=607}{Spectrum} \\
       2 & J000431.92-011411.8 & 0.0888 & 19.93 & 17.96 & 17.13 & 16.69 & 16.39 & -21.33 & 10.47 & 2.29 & 1.14 & 1.97 & 2.27 & 1.90 & 1.55 & \href{http://skyserver.sdss.org/dr9/en/tools/explore/obj.asp?ra=1.1330&dec=-1.2366}{Object} & \href{http://dr9.sdss3.org/spectrumDetail?plateid=388&mjd=51793&fiber=282}{Spectrum} \\
       3 & J001027.38-104341.9 & 0.1329 & 19.42 & 17.81 & 17.07 & 16.73 & 16.46 & -22.51 & 10.78 & 2.03 & 1.49 & 2.30 & 2.12 & 1.62 & 1.19 & \href{http://skyserver.sdss.org/dr9/en/tools/explore/obj.asp?ra=2.6141&dec=-10.7283}{Object} & \href{http://dr9.sdss3.org/spectrumDetail?plateid=651&mjd=52141&fiber=92}{Spectrum} \\
       4 & J001145.22-005430.6 & 0.0479 & 18.73 & 16.94 & 16.28 & 15.93 & 15.63 & -20.87 & 10.13 & 2.60 & 1.45 & 1.93 & 2.82 & 2.50 & 2.18 & \href{http://skyserver.sdss.org/dr9/en/tools/explore/obj.asp?ra=2.9384&dec=-0.9085}{Object} & \href{http://dr9.sdss3.org/spectrumDetail?plateid=388&mjd=51793&fiber=11}{Spectrum} \\
       5 & J001556.50+141151.0 & 0.0834 & 20.03 & 18.18 & 17.42 & 16.96 & 16.68 & -20.75 & 10.23 & 2.02 & 1.17 & 1.31 & 1.53 & 1.40 & 0.86 & \href{http://skyserver.sdss.org/dr9/en/tools/explore/obj.asp?ra=3.9854&dec=14.1975}{Object} & \href{http://dr9.sdss3.org/spectrumDetail?plateid=752&mjd=52251&fiber=58}{Spectrum} \\
       6 & J001717.14+140040.7 & 0.1442 & 19.29 & 17.67 & 16.81 & 16.39 & 16.09 & -22.94 & 11.02 & 2.40 & 1.63 & 2.02 & 2.24 & 1.73 & 1.26 & \href{http://skyserver.sdss.org/dr9/en/tools/explore/obj.asp?ra=4.3214&dec=14.0113}{Object} & \href{http://dr9.sdss3.org/spectrumDetail?plateid=752&mjd=52251&fiber=14}{Spectrum} \\
       7 & J002928.97+143342.8 & 0.1431 & 19.59 & 18.31 & 17.45 & 17.01 & 16.73 & -22.81 & 10.80 & 2.47 & 1.79 & 2.34 & 2.30 & 1.68 & 1.10 & \href{http://skyserver.sdss.org/dr9/en/tools/explore/obj.asp?ra=7.3707&dec=14.5619}{Object} & \href{http://dr9.sdss3.org/spectrumDetail?plateid=417&mjd=51821&fiber=268}{Spectrum} \\
       8 & J003002.90-005306.7 & 0.0600 & 19.13 & 17.48 & 16.78 & 16.45 & 16.21 & -20.62 & 10.06 & 2.08 & 1.24 & 1.52 & 2.02 & 1.58 & 1.19 & \href{http://skyserver.sdss.org/dr9/en/tools/explore/obj.asp?ra=7.5121&dec=-0.8852}{Object} & \href{http://dr9.sdss3.org/spectrumDetail?plateid=391&mjd=51782&fiber=49}{Spectrum} \\
       9 & J003209.48-091332.9 & 0.1676 & 20.02 & 18.32 & 17.49 & 17.09 & 16.81 & -22.32 & 10.83 & 2.19 & 1.51 & 1.53 & 2.08 & 1.59 & 1.18 & \href{http://skyserver.sdss.org/dr9/en/tools/explore/obj.asp?ra=8.0395&dec=-9.2258}{Object} & \href{http://dr9.sdss3.org/spectrumDetail?plateid=654&mjd=52146&fiber=461}{Spectrum} \\
      10 & J003248.96-100043.9 & 0.0131 & 16.48 & 15.29 & 15.01 & 14.90 & 14.75 & -19.03 & 9.124 & 2.03 & 1.45 & 1.67 & 1.38 & 1.51 & 0.81 & \href{http://skyserver.sdss.org/dr9/en/tools/explore/obj.asp?ra=8.2040&dec=-10.0122}{Object} & \href{http://dr9.sdss3.org/spectrumDetail?plateid=654&mjd=52146&fiber=499}{Spectrum} \\
      11 & J003402.78-094219.1 & 0.0125 & 14.45 & 12.63 & 11.76 & 11.31 & 10.97 & -22.02 & 10.77 & 2.42 & 1.77 & 2.06 & 2.56 & 2.20 & 1.70 & \href{http://skyserver.sdss.org/dr9/en/tools/explore/obj.asp?ra=8.5116&dec=-9.7053}{Object} & \href{http://dr9.sdss3.org/spectrumDetail?plateid=654&mjd=52146&fiber=519}{Spectrum} \\
      12 & J003409.94-001718.6 & 0.0580 & 18.91 & 17.73 & 17.47 & 17.31 & 17.33 & -19.83 & 9.481 & 2.17 & 1.64 & 1.80 & 1.48 & 1.58 & 0.96 & \href{http://skyserver.sdss.org/dr9/en/tools/explore/obj.asp?ra=8.5414&dec=-0.2885}{Object} & \href{http://dr9.sdss3.org/spectrumDetail?plateid=392&mjd=51793&fiber=278}{Spectrum} \\
      13 & J003707.82+002436.4 & 0.0807 & 19.35 & 17.68 & 17.01 & 16.66 & 16.45 & -21.03 & 10.25 & 2.11 & 1.33 & 1.77 & 2.24 & 1.80 & 1.50 & \href{http://skyserver.sdss.org/dr9/en/tools/explore/obj.asp?ra=9.2826&dec=0.4101}{Object} & \href{http://dr9.sdss3.org/spectrumDetail?plateid=392&mjd=51793&fiber=487}{Spectrum} \\
      14 & J003921.65+002219.9 & 0.0827 & 20.15 & 18.46 & 17.62 & 17.16 & 16.79 & -20.56 & 10.13 & 1.94 & 1.23 & 1.22 & 1.82 & 1.58 & 0.86 & \href{http://skyserver.sdss.org/dr9/en/tools/explore/obj.asp?ra=9.8402&dec=0.3722}{Object} & \href{http://dr9.sdss3.org/spectrumDetail?plateid=392&mjd=51793&fiber=593}{Spectrum} \\
      15 & J004136.31+004638.3 & 0.1019 & 19.03 & 17.78 & 17.13 & 16.73 & 16.51 & -21.41 & 10.34 & 2.26 & 1.63 & 2.04 & 1.97 & 1.62 & 0.91 & \href{http://skyserver.sdss.org/dr9/en/tools/explore/obj.asp?ra=10.4013&dec=0.7773}{Object} & \href{http://dr9.sdss3.org/spectrumDetail?plateid=393&mjd=51794&fiber=385}{Spectrum} \\

\hline \hline
\end{tabular}
\label{tab:catalog} } \\
\raggedright {\footnotesize
The first 15 SPOG* candidates. Column (1): SPOG catalog index. (2): SDSS IAU name. Column (3): SDSS redshift. Columns (4-8): SDSS {\em u\,g\,r\,i\,z} dereddened model magnitudes from DR9. Column (9): Absolute {\em r} magnitude. Column (10): Log of the mass (calculated in \S\ref{sec:colors}). Columns (11-16): Logarithm of the narrow line observed fluxes from OSSY (in 10$^{-17}$ ergs~s$^{-1}$~cm$^{-2}$). Column (17): SDSS DR9 Explorer Page link for object. Column (18): SDSS DR9 Spectral Explorer link for object. \label{tab:catalog}
} 
\vspace{1mm}
\end{table}

30,225 (19\%) of the ELGs catalog fall within these shock model boundaries across all three line diagnostic diagrams. We note that there is significant overlap of the shock boundaries with star formation (SF), composites, Seyferts and LINERs, and therefore we do not expect the majority of galaxies that fall within the shock boundaries to be dominated by shocks. Furthermore, we do not expect galaxies to harbor a single source of ionization, rather that they could harbor shocks as well as additional sources of ionization. Seyferts with log(\oiii/H$\beta$)\,$>$\,1.03 can be distinguished from shocks, however even a large portion of quasars with poststarburst signatures fall below the traditional Seyfert cutoff of log(\oiii/H$\beta$)\,$>$\,1.03  and therefore also within the shock boundaries \citep{cales+13}.

\begin{figure*}[t]
\centering
\includegraphics[width=\textwidth]{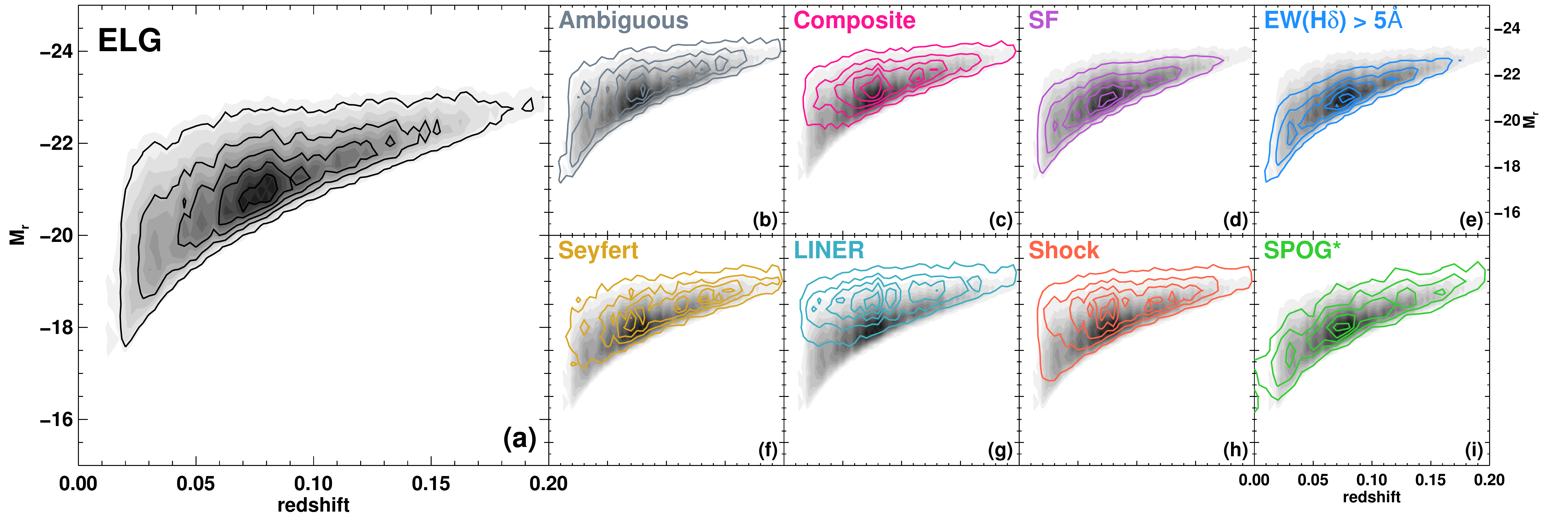}
\caption{\footnotesize SDSS absolute {\em r}-band magnitude versus redshift for the ELG (a; black, grayscale) overplotted with the distribution of objects that are spectrally classified as ambiguous (b:\,gray), composite (c:\,pink), SF (d:\,purple), those meeting the \ewhd\ criterion (e:\,blue), Seyferts (f:\,yellow), LINERs (g:\,turquoise), objects within the shock boundaries (h:\,red), and SPOGs* (i:\,green). Contours for all subsamples are in increments of 10\% from the maximum in each distribution.  All objects are well represented in redshift space.
Composites, Seyferts, and LINERs tend to have large M$_r$ and make up a higher fraction of the ELG at the higher end of our redshift range ({\em z}\,=\,0.2). SPOGs* typically have lower M$_r$, particularly at low z. 
}
\label{fig:Mr_z}
\end{figure*}

LINER emission accounts for the majority of phase space defined within the shock boundaries, and the largest portion (36\%) of shock candidates from the ELG sample. Seyferts account for the next largest subset (14\%), followed by composites (6\%), and SF (4\%).  The remaining 40\% of shock candidates have ambiguous line diagnostic classifications (fairly consistent with the 43\% of shock models that produce ambiguous classifications; \citealt{dopita+05,allen+08,rich+11}).  The requirement that shock candidates must be found within the shock boundaries within all three line diagnostic diagrams mitigates the risk of significant contamination from the SF region, which comprises the majority of the ELG sample.

\subsection{Prevalence of Star Forming Contaminants}
\label{sec:sf_contaminants}

A total of 4,508 (3\%) objects have line diagnostics consistent with shocks and Balmer absorption indicative of a strong A-star population (\ewhd). We note that a large number (3,441, 76\%) of these systems have narrow line emission diagnostics matching models of star formation. 
For this reason, to define a sample of pure poststarburst, shocked candidates, we remove all galaxies with \ewhd\ that consistently fall within the Composite or SF region of the line diagnostic diagrams \citep{kewley+06} for all three diagnostic diagrams.  While it is possible that we are missing very interesting galaxies that fall in the region where star formation and shocks overlap, the data in-hand do not allow us to distinguish between the true shocked galaxies and star-forming galaxies.

The properties of the shock models that fall within the SF region of the line diagnostic diagrams tend to be those with low metallicities and higher densities, and thus are missed from our shock definition. There is also the possibility of contamination of ``red bulge'' galaxies.  Below {\em z}\,=\,0.016, the 3$''$ SDSS fiber covers less than 1\,kpc diameter.  If the 3$''$ SDSS fiber only covers the nucleus of a bulge-dominated spiral, it could be classified as a shock candidate, despite the fact that the integrated emission from the galaxy would be clearly classified as star-forming.  In 55 SPOGs*, less than 10\% of the optical {\em i}-band light went down the fiber (calculated by taking the ratio of the fiber and Petrosian magnitudes). Many of these were at very low redshift ({\em z}\,$<$\,0.01), but others appeared to be due to offset fiber positions. We inspected these objects; while some were contaminants, many others showed signs of being real SPOGs. For completeness, we did not remove these 55 SPOGs* from the catalog.

To determine the contamination fraction of ``red bulge'' galaxies (galaxies that contained extended star-forming disks and quiescent nuclei) that were not covered by the 3$''$ SDSS fiber, we manually inspected all identified SPOGs* with redshifts below {\em z}\,$<$\,0.03 (a total of 109 objects), identifying galaxies with clear signatures of star formation outside of the nucleus. Most SPOG candidates (81\%) below {\em z}\,$<$\,0.01 exhibited signs of extended star formation in their outer disks. The contamination fraction reduces rapidly with increasing redshift increments (34\% at 0.025\,$\leq$\,{\em z}\,$<$\,0.03, down to 4\% at 0.04\,$\leq$\,{\em z}\,$<$\,0.06). The SPOG selection has strongly redshift-dependent contamination effects, although we did not remove these objects from the survey to remain complete in our criterion selection. We recommend those selecting the lowest redshift SPOGs* to carefully inspect those objects. A more detailed discussion of biases within the ELG and SPOG* sample can be found in Appendix~\ref{app:bias}. A more detailed description of the morphologies of SPOGs is beyond the scope of this paper.

\subsection{The Shocked Poststarburst Galaxy (SPOG) Sample}

We define the Shocked Poststarburst Galaxy (SPOG) sample as objects from the ELG catalog that simultaneously meet the \ewhd\ and shocked criteria, while also not falling consistently inside the SF and composite regions of all three line diagnostic diagrams.  1,067 objects (0.7\% of the 159,387 ELGs) meet this criterion. Given that location within the shock boundaries does not confirm the presence of shocks, we identify these objects as candidates, or SPOGs*. We visually inspected the thumbnails for each SPOG* to ensure that no galaxies included duplicate spectra, and found that in all cases, each SPOG-identified spectrum represented a single object.

\begin{figure*}[t]
\centering
\includegraphics[width=\textwidth]{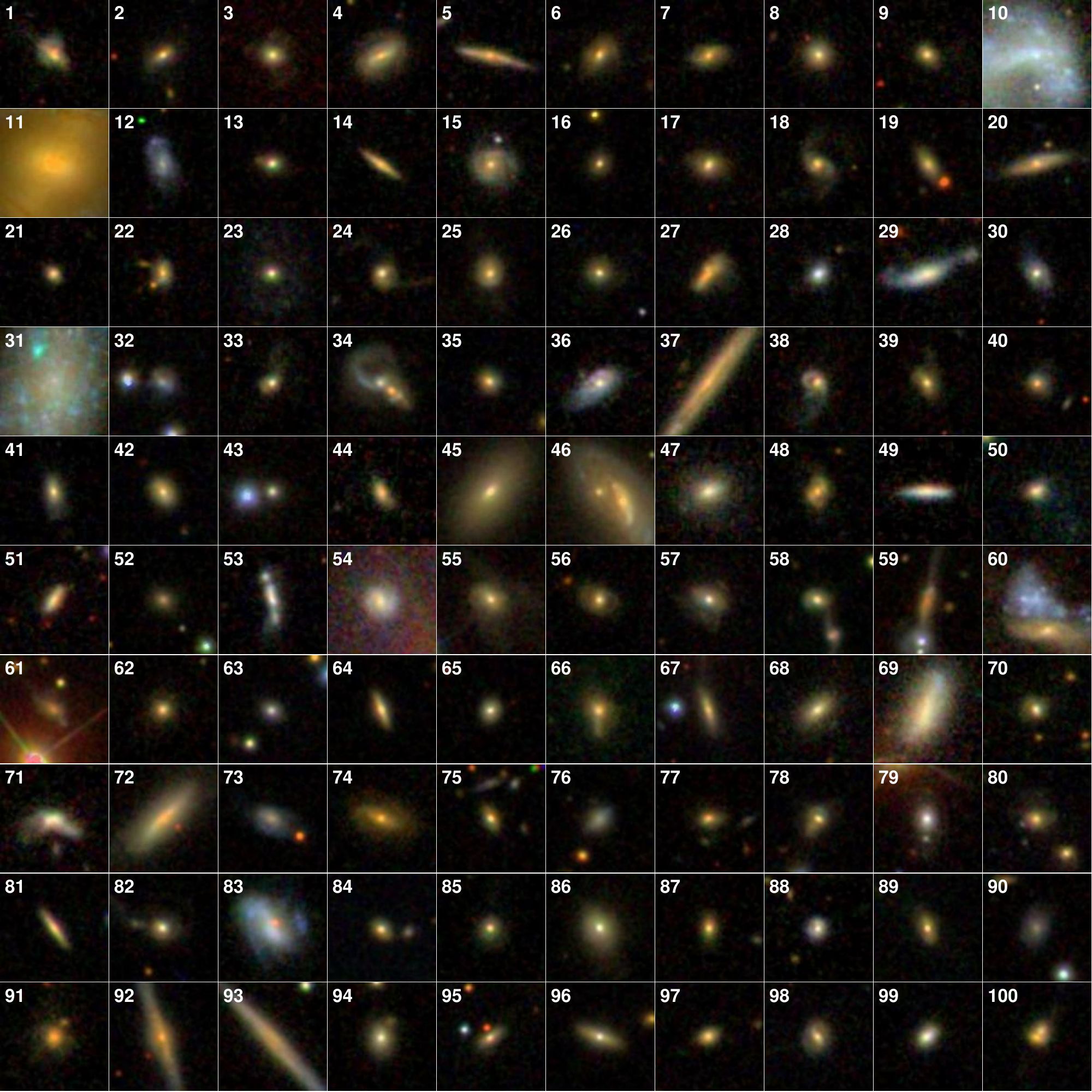}
\caption{\footnotesize SDSS {\em g\,r\,i} images of the first 100 SPOGs*, laid out in order of right ascension. Each thumbnail was pulled from SDSS DR12 \citep{sdssdr12}, with fields of view of 30$''$ in all cases. The 3$''$ SDSS fiber is located at the center of the image. All other thumbnails are available in the online material, and are representative of the morphologies seen throughout the rest of the survey. Thumbnails of all SPOGs can be found at \href{http://www.spogs.org}{http://www.spogs.org}}
\label{fig:spogthumbsstart}
\end{figure*}

\begin{figure*}[t]
\centering
\includegraphics[width=0.99\textwidth]{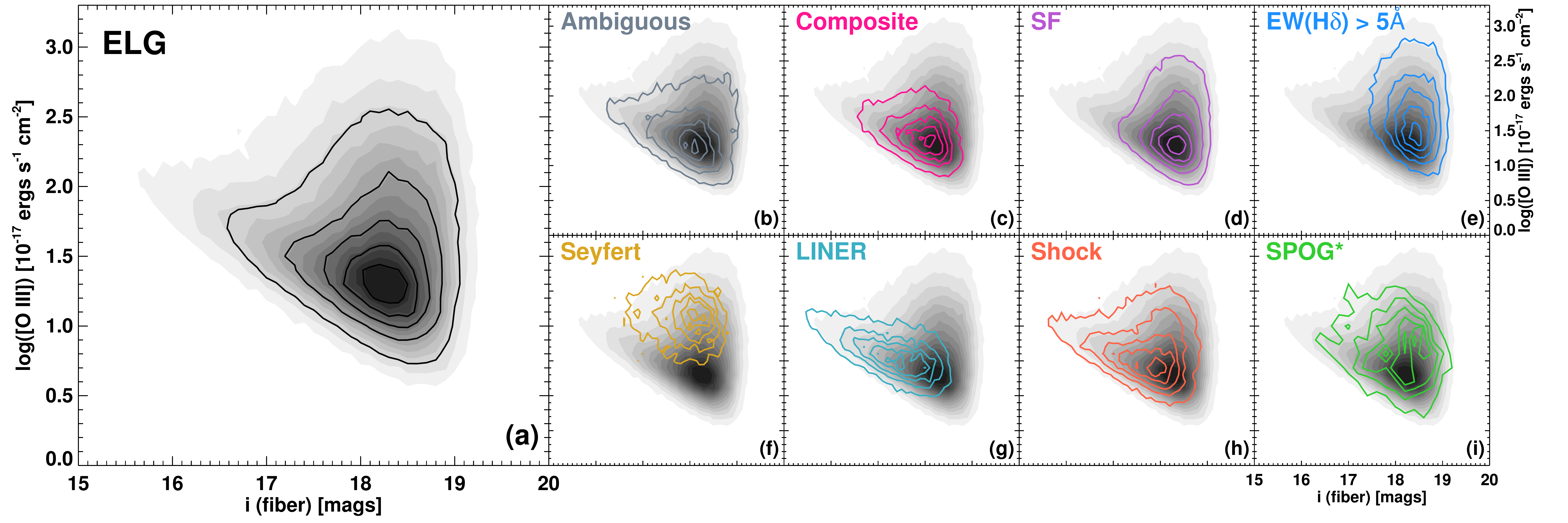}
\caption{\footnotesize \oiii\ flux versus apparent {\em i}-band fiber magnitude for the ELG (a; black, also grayscale) overplotted with the distribution of objects that are spectrally classifiied as ambiguous (b:\,gray), composite (c:\,pink), SF (d:\,purple), those meeting the \ewhd\ criterion (e:\,blue), Seyferts (f:\,yellow), LINERs (g:\,turquoise), objects within the shock boundaries (h:\,red), and SPOGs* (t:\,green).  
Contours for all subsamples are in increments of 10\% of the maximum in each distribution.  The ELG seems to show two branches, with one in which \oiii\ emission excited proportional to the stellar mass, most apparent in LINERs, but also present in ambiguous classifications, composite objects and shocks.}
\label{fig:spogs_ioiii}
\end{figure*}

Figure~\ref{fig:spog_bpt} shows the \oiii/H$\beta$ versus \nii/H$\alpha$, \sii/H$\alpha$, and \oi/H$\alpha$ line diagnostic diagrams \citep{bpt,veilleux+87} for the entire ELG sample (grayscale), with the SPOGs* plotted as green points.  The shock boundaries are shown as purple lines, and the different line ratio population divisions \citep{kauffmann+03,kewley+06} are shown as black lines. SPOGs* span the range defined by the shock boundaries in \S\ref{sec:shocks} and contain nearly equal contributions from objects classed as LINERs (186; 17$\pm$1.1\%) and Seyferts (194; 18$\pm$1.2\%) with the remainder having ambiguous classifications. 

To test how significantly aperture bias affects the catalog, we plot the distribution of SDSS absolute {\em r}-band magnitude (M$_r$, corrected for Galactic extinction) versus {\em z} in Figure \ref{fig:Mr_z} for the objects that have been spectrally classified as ambiguous, composite, SF, Seyfert, and LINER; objects meeting the \ewhd\ criteria; those within the shock boundaries; and SPOGs*. In M$_r$--{\em z} space, SPOGs* follow similar trends to star-forming galaxies and \ewhd\ objects, rather than LINER or shock-boundary objects, populating the low $M_r$ region at low redshift (further discussion can be found in \S\ref{app:bias}). 

This SPOG* selection criteria suite was also used in \citet{a14_irtz}, who show that SPOGs* as a population are likely to be found in the green valley and have Wide-Field Infrared Survey Explorer ({\em WISE}; \citealt{wise}) infrared colors consistent with a transitioning population.  Given that galaxy color was not a part of the criteria used to identify SPOGs*, this result is reassuring, in that, by using the SPOGs* criteria, we have indeed identified a population that appears to be quenching.

Figure \ref{fig:spogthumbsstart} presents 30$''$ {\em g\,r\,i} SDSS cut-outs of the first 100 of 1,067 SPOGs* in our sample\footnote{All thumbnails can be accessed through the online material, as well as via \href{http://www.spogs.org}{http://www.spogs.org}}, and Table~\ref{tab:catalog} presents the SPOG identifications, redshifts, {\em u\,g\,r\,i\,z} de-reddened magnitudes from SDSS DR9 \citep{sdssdr9}, emission line fluxes from OSSY \citep{ossy}, M$_r$, {\em i}-band determined masses, and links to the SDSS explorer page. The catalog of SPOGs* is available through the NASA/IPAC Extragalactic Database (NED)\footnote{\href{http://ned.ipac.caltech.edu/}{http://ned.ipac.caltech.edu/}}, this journal article, and the SPOGs \href{http://www.spogs.org}{website}.

\section{Discussion}
\label{sec:disc}
\subsection{Investigating ionization mechanisms of SPOGs*}

As discussed in \S\ref{sec:shocks}, there are many mechanisms capable of creating emission line ratios consistent with shocks, such as star formation, AGN activity, and photoionization from post-asymptotic giant branch (post-AGB) stars. In \S\ref{sec:sf_contaminants}, we identified a subset of galaxies with narrow-line diagnostics within the shock boundaries, but indicative of star formation and thus mitigate contamination by omitting these systems from SPOGs*. For the remaining SPOGs*, we examine the likely ionization mechanisms and consider future observations to distinguish between them. 

\begin{figure*}[t]
\centering
\includegraphics[width=0.99\textwidth]{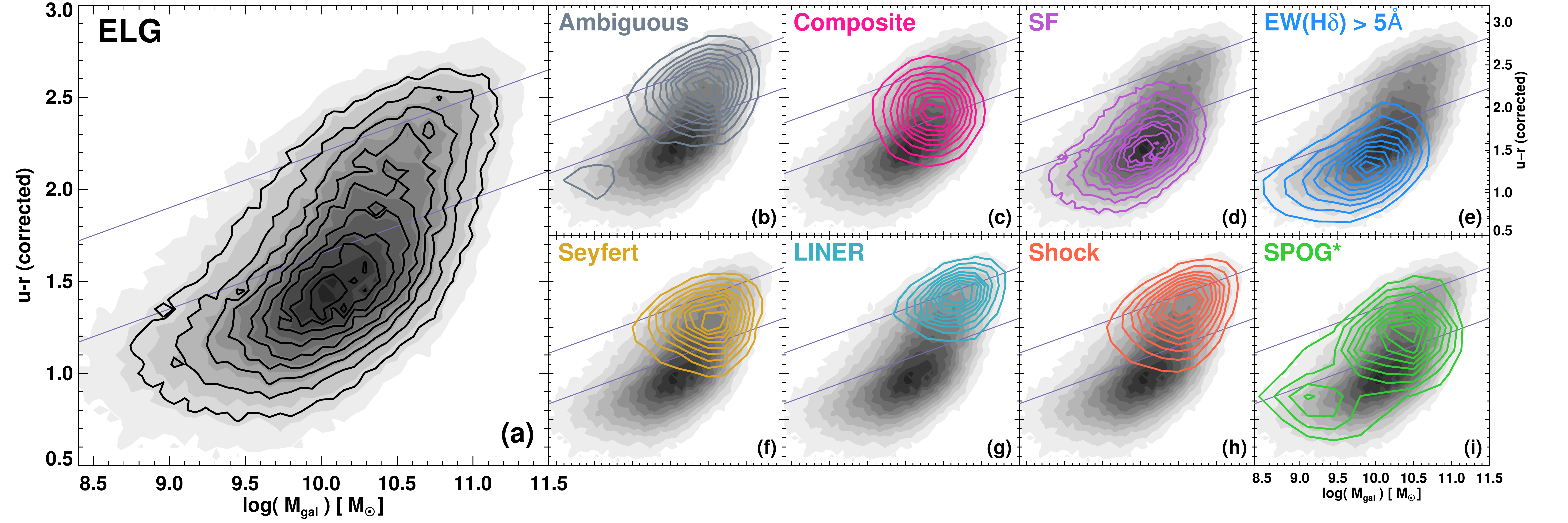}
\caption{\footnotesize Corrected {\em u--r} color-mass diagram (see \S\ref{sec:colors}) calculated for the ELG sample (a; black, also grayscale) overlaid with distributions of objects spectrally classified as ambiguous (b:\,gray), composite (c:\,pink), SF (d:\,purple), meeting the \ewhd\ criterion (e:\,blue), Seyferts (f:\,yellow), LINERs (g:\,turquoise), objects within the shock boundaries (h:\,red) and SPOGs* (i:\,green). Contours for all subsamples are 10\% of the maximum.  The indigo lines represent the green valley defined by \citet{schawinski+14}. }
\label{fig:spogs_cmd}
\end{figure*}

If the ionized gas originates from post-AGB stars, the narrow line emission (including the \oiii) should correlate with the stellar mass \citep{capetti+baldi11,yan+12}. Figure~\ref{fig:spogs_ioiii} shows the distributions of \oiii\ versus SDSS {\em i}-band fiber magnitudes of the ELG and each of our subsamples. ELGs show two distinct branches, one shows a clear correlation between  \oiii\  and {\em i}-band flux, and the other branch at fainter {\em i}-band magnitude is uncorrelated with \oiii. 
LINERs show the strongest relationship between {\em i}-band and \oiii, consistent with the hypothesis that LINER emission mainly originates {in post-AGB stars} in massive, early-type systems (\citealt{yan+06}; also confirmed by the red colors in Fig.\,\ref{fig:spogs_cmd}).  This relationship is also present in the ambiguous subsample, composites, and shocks, but is weak or absent from the SPOGs* distribution.  SPOGs* look more like SF or Seyferts than LINERs or shocks, consistent with the idea that the ionized gas in SPOGs* are not primarily excited by post-AGB stars.

It is likely that a non-negligible number of SPOGs* contain AGNs (consistent with what is found in NGC\,1266; \citealt{alatalo+11,davis+12}), but it is unclear the total fraction of SPOGs* in which an AGN photoionization is the dominant mechanism exciting the ionized gas.  The fact that a minority (18\%) of SPOGs* exhibit Seyfert-like line ratios means that our contamination from moderate (Seyfert) luminosity AGNs does not overwhelm the SPOGs* sample.  Low-luminosity AGNs (LLAGNs) could also mimic the shock emission of the SPOGs* sample.  LLAGNs are able to exhibit both LINER-like and ambiguous emission line ratios, though the emission would be isolated to the nucleus. LLAGN therefore could be a contamination source, especially at low-{\em z} where the 3$''$ SDSS fiber is only able to probe the ionization environment in the nucleus of the galaxy.  As the SDSS fiber subtends larger area on the galaxy, it probes larger physical regions, possibly finding shock-like emission, which is usually extended. A more detailed discussion of how the line diagnostics and SPOG* detections change over redshift and mass can be found in Appendix \ref{app:bias}.


Integral Field Spectroscopy (IFS) shows promise at identifying shocks. A recent IFS study of NGC\,7130 suggested that the two \nii/H$\alpha$ peaks (which lacks coincident peaks of emission in the near and mid-infrared) are energized by shocks tracing an outflowing wind, possibly by an AGN or star formation \citep{davies+14}.  IFS studies of extended LINER-like emission revealed the presence of extended shock excitation in nearby major galaxy mergers \citep{monreal-ibero+06,monreal-ibero+10,rich+10,rich+11,rich+14,rich+15}. Extended shocks are often indicative of galactic wind-driven feedback and are easily discernible with IFS, but may appear as simple AGN or composite-like emission in spatially unresolved spectroscopy \citep{rich+14}. 

Large IFS studies such as the Calar Alto Legacy Integral Field Area (CALIFA), the Sydney AAO Multi-Object Integral (SAMI) Field Spectrograph, and the Mapping Nearby Galaxies at APO (MaNGA) surveys are observing thousands of galaxies with IFS \citep{sanchez+12,croom+12,bundy+15}. These studies have already revealed the presence of shocks and winds in more ``normal'' nearby galaxies (e.g. \citealt{kehrig+12,fogarty+12,ho+14,ho+16}).  Studying SPOGs* with IFS will not only reveal interesting substructure and 2{\sc d} kinematics of the emission present but will be able to remove a possible contaminant to the sample: AGNs.  SPOGs* with emission that is excited by shocks will appear extended when observed with an IFS, while a LLAGN will manifest as a point source consistent with the nucleus.

SPOGs* have optical emission-lines consistent with shocks, but optical diagnostic ratios alone are degenerate with other processes. Near-IR emission lines such as [Fe\,{\sc ii}] and ro-vibrationally excited H$_2$ are able to break this degeneracy. Shock excitation manifests itself in large [Fe~{\sc ii}]1.64/Br$\gamma$ emission line ratios \citep{mouri+00} and strong H$_2$ lines. 
Thus, [Fe\,{\sc ii}] line detection in conjunction with the H$_2$ lines will reveal differences in excitation and kinematics away from the nuclei (see \citealt{mouri+00}).  The two main excitation routes for ro-vibrational molecular hydrogen emission are UV florescence and collisions in hot gas in shocks \citep{shull+82,black+87}, and these can be distinguished by looking at the relative strength of the ro-vibrational line fluxes, 2-1S(1)/1-0S(1); typically 0.1 for T\,=\,2000\,K shocks, and $\sim$\,0.5 in UV-pumped star-forming regions. Detection of multiple transitions of ro-vibrational H$_2$ will be able to unambiguously identify the presence of shocks in SPOGs*, and if the spectra are taken using a longslit spectrograph or an IFS, extended shocks can be mapped, searching for emission similar to the extended winds seen in NGC\,1266 \citep{alatalo+11} or high redshift quasars \citep{nesvadba+08}.  The first results of near-IR observations of ro-vibrationally excited H$_2$ detections in SPOGs* will be presented in an upcoming paper (Alatalo et al. 2016, in preparation).

\begin{figure}[b]
\includegraphics[width=0.48\textwidth,clip,trim=0cm 0.3cm 0cm 0.6cm]{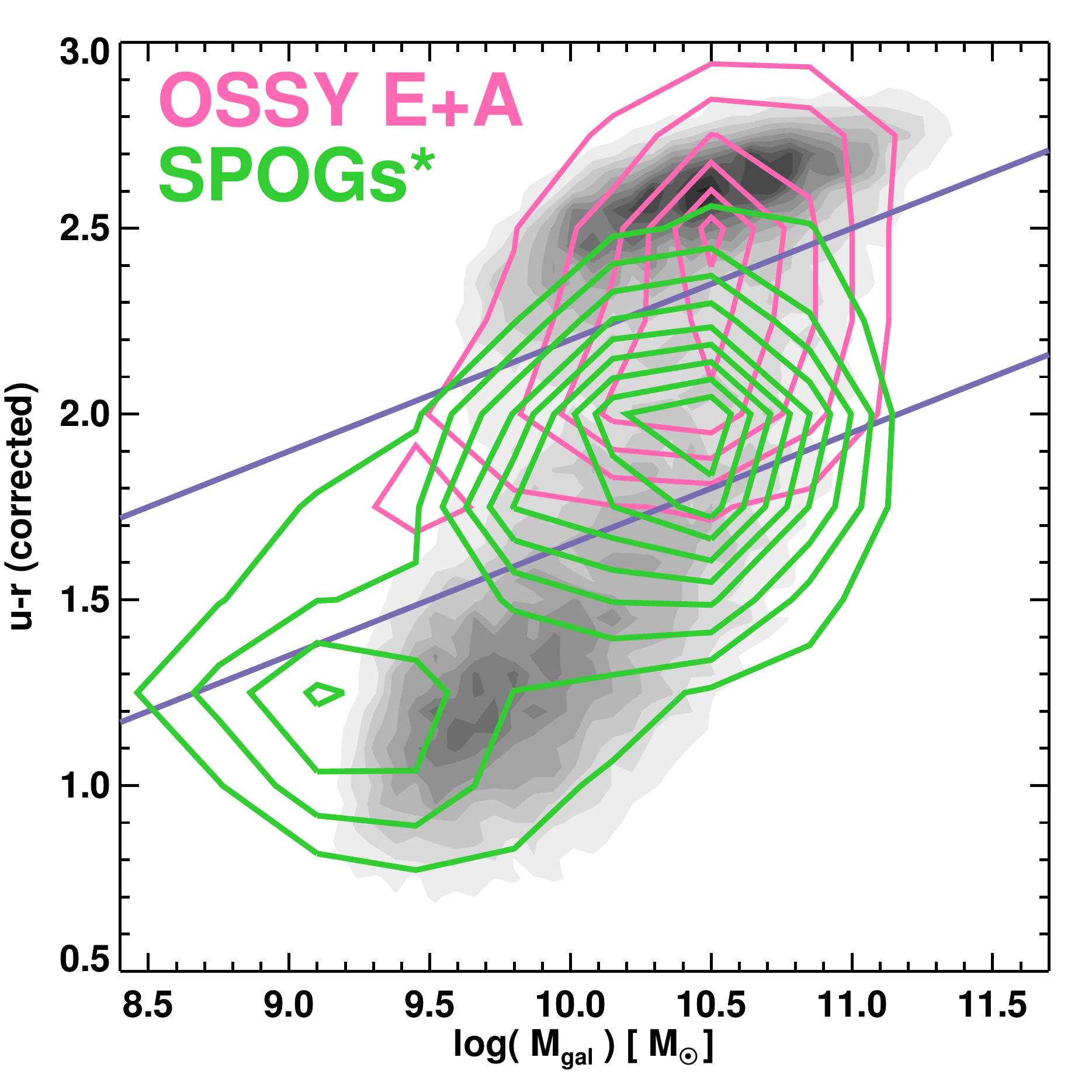}
\caption{\footnotesize The optical color-mass distributions for the Galaxy Zoo sample (\citealt{schawinski+14}; grayscale), compared to SPOGs* (green contours) and ``E+A'' (criteria from \citealt{goto07}) galaxies from the OSSY catalog that meet the continuum S/N requirements of the ELG catalog (pink contours). The OSSY ``E+A'' galaxies are found in the optical green valley (as expected; \citealt{dressler+gunn83}) with a wing in the red sequence. Poststarburst galaxies are also (on average) more massive than SPOGs* (though it is likely this is due to the signal to noise cuts applied to the OSSY E+A sample). The green valley definition from \citet{schawinski+14} is bounded by indigo lines. The SPOG* contours confirm that SPOGs* on average are bluer than poststarburst galaxies, including in comparable mass bins.}
\label{fig:psb_cmd}
\end{figure}

\subsection{Are SPOGs* entering or leaving the ``green valley''?}

Figure\,\ref{fig:spogs_cmd} presents {\em u--r} color-mass diagram for the ELG and SPOG* samples. The ELG catalog, shown in the large panel to the left and as the grayscale contours underlying the subsamples, primarily occupies the blue cloud.  Contours of each of the subsamples described above are overplotted in the small panels. Composites tend to be slightly bluer than Seyferts, and are much more likely to be found in the green valley than pure star-forming objects.  Objects classified as ambiguous contain two optical color peaks: one similar to Seyferts that might be galaxies lying along the Seyfert-LINER mixing line \citep{rich+14}, and one at low-mass low-redshift that might be low metallicity dwarfs (discussed in Appendix~\ref{app:bias}).  Objects that fall within shock boundaries are most likely to be near the red sequence, similar to LINERs.  \ewhd\ objects have the bluest peak of all subsamples represented here, which is probably due to their likelihood to harbor narrow-line ratios typical of star forming regions, and their high incidence of A-stars, which peak in blue optical light. SPOGs* have a high fraction of objects in the green valley, but with a bluer peak than composites or Seyferts, thus they likely have a larger population of young stars. SPOGs* (according to their narrow-line diagnostics) are not actively forming stars, and thus are not as blue as the star-forming and \ewhd\ objects.

The location of SPOGs* on the blue edge of the green valley (Figure\,\ref{fig:spogs_cmd}i) indicates that they might be recent arrivals to the green valley.  As described in \citet{a14_irtz}, SPOGs* are also found to have {\em WISE} colors consistent with entering the transition zone rather than leaving. Figure\,\ref{fig:psb_cmd} compares the {\em u--r} colors of SPOGs* to the colors\footnote{{\em k}-corrected, dereddened, extinction corrected colors derived from the SDSS modelmags} of OSSY galaxies that pass the continuum S/N cuts used for the ELG sample (\S\ref{sec:catalog}), but with an additional ``E+A'' criterion of \citet{goto07}\footnote{H$\delta$ absorption: \ewhd, absense of emission: EW([O\,{\sc ii}])\,$<\,$2.5\AA, EW(H$\alpha$)\,$<$\,3\,\AA} applied. The OSSY E+A objects on average have green colors, with a wing extending to the red sequence.
The fact that the SPOGs* peak is bluer than the E+A peak (objects that are known to be transitioning), but still consistent with the optical green valley is a promising sign that we have identified the transforming population that we are searching for. When studied in detail, SPOGs* might be able to elucidate the behaviors of galaxies at earlier stages of transition, especially given that green valley and poststarburst galaxies appear to show signs of having morphologically transitioned \citep{yang+04,yang+08,wong+12}.  In order to detail the evolution of SPOGs*, analyses of the UV properties, spectral energy distributions and stellar population synthesis modeling are underway (Lanz et al. 2016, in preparation).


It is also possible that SPOGs* are not late-type galaxies entering the green valley from the blue sequence, but rather early-type galaxies that have been replenished in molecular gas and entered the green valley from the red sequence. \citet{dressler+13} presented a large sample of galaxies inside and outside clusters between 0.3\,$<$\,{\em z}\,$<$\,0.5. The authors found that galaxies with poststarburst stellar signatures had a mass distribution consistent with that of the early-type galaxies, whereas the mass distribution of starbursting galaxies was consistent with the mass distribution of late-type galaxies, independent of environment. The authors went on to suggest that this mismatch argued for different origins for poststarburst and starburst galaxies, with the majority of poststarburst galaxies being replenished early-types. Thus, as with the poststarburst galaxies from \citet{dressler+13}, it is possible that the origin of SPOGs* is not quenching late-type galaxies, but rather early-type galaxies with replenished gas reservoirs. In this case, the difference between SPOGs* and poststarbursts could be either a larger accreted gas reservoir or objects caught earlier in the process of replenishing. A detailed mass distribution, similar to what was done in \citet{dressler+13} will be required to determine if replenished early-types are a major contributor to the SPOG phase.

\subsection{The SPOG Lifetime}
Poststarburst stellar populations occur when current star formation terminates abruptly ($<$\,100\,Myr) and remains dormant for several 100\,Myrs. Typical spectra display Balmer jumps and high order Balmer absorption lines common of A-type stars. The detectability of these spectral signatures is set to first order by the main sequence lifetimes of A-stars ($\sim$\,1\,Gyr; \citealt{falkenberg+09}), but in most cases is only detectable for 0.1\,--\,0.3\,Gyr \citep{snyder+11}. Thus, these galaxies are thought to be in transition between actively star-forming late-type galaxies and passive early-type galaxies observable on a timescale of 100s of Myrs. This is consistent with the notion that early-type galaxies move rapidly across the green valley from blue cloud to red sequence in several 100\,Myr to $\sim$\,1\,Gyr, after their morphologies are transformed from disk to spheroid and star formation is quenched rapidly \citep{schawinski+14}. 

The 1,067 SPOG candidates represent 0.2\% of the OSSY sample that meet the continuum S/N requirements (591,627 objects), and 0.7\% of the ELG sample. Catalogs of poststarburst galaxies usually number in the several hundreds and tend to account for 0.2\% of galaxies \citep{zabludoff+96, quintero+04, goto07, pattarakijwanich+14}. There were 694 (0.1\%) objects in the OSSY continuum sample that met the E+A criteria of \citet{goto07}, within a factor of two of the number of SPOGs* we identify. In both the cases of OSSY E+A and SPOG samples, selection biases remain. In the case of SPOGs*, our line intensity criterion (the significant detection of all diagnostic lines) likely removes SPOGs* from the survey, so 0.2\% is a lower limit. Similarly, poststarburst galaxies include only objects with weak or absent nebular line ([O\,{\sc ii}] or H$\alpha$) emission, despite there being multiple non-starforming mechanisms capable of exciting them. In this case, because the SPOG and ``E+A'' criterion sample a nearly mutually exclusive parameter space, it is possible that we can add SPOGs* to the overall poststarburst/post-transition galaxy sample. Our new method of selecting post-transition galaxies, while allowing for them to harbor powerful sources of ionization from AGN, LINERs, and/or shocks, provides a sizable, heretofore neglected, fraction of the total transitioning population. 

If we assume that the fraction of SPOGs* among the OSSY continuum sample roughly represents the fraction of SPOGs* in the local universe, we can derive a rough estimate of the amount of time that galaxies spend in a SPOG phase. Assuming that all galaxies go through a single SPOG phase over a Hubble time, we find an average SPOG timescale of $t_{\rm SPOG}$\,$\approx$\,20\,Myr. Given the selection biases in place (including the requirement of finding strong \oi\ emission), this is likely a lower limit on the timescale. Assuming $t_{\rm SPOG}$ represents a lifetime of a SPOG episode, the fact that it roughly agrees with the dissipation time of shocks ($\sim$10-100\,Myr; \citealt{guillard+09,lesaffre+13}), suggests that the SPOG lifetime might be driven by the timescale that shocks are illuminating the galaxy, rather than the timescale over which intermediate-aged stars can be detected. This is likely an additional reason that the poststarburst lifetime does not match the timescale of A-stars: that non-star forming ionization mechanisms might be producing nebular line emission in the galaxy after quenching, causing the ``E+A'' survey to miss transitioning objects. 

In the context of merger driven evolutionary scenarios, rapid bursts of star formation ($\tau_{\rm quench}$ $\lesssim$100 Myr) happen early in the merger and coincide with the close passages of the two nuclei \citep{dimatteo+05, hopkins+08, vanwassenhove+12, stickley+14}. While the merger and its signatures can last 1-2 Gyr \citep{dimatteo+05}, each successive pass of the nuclei occur on timescales of several 100 Myrs \citep{stickley+14}, and final coalescence of the nuclei is quick ($\lesssim$100 Myr; \citealt{lanz+14}). Mergers are effective at triggering shocks \citep{rich+11,rich+14,rich+15, soto+12, inami+13}; because the merger rate the nearby Universe \citep{darg+10,lotz+11} agrees with the SPOG* fraction, so it is possible that mergers undergo a SPOG phase. Given that the starburst timescale within mergers is longer than the shock dissipation time, signs of the SPOG phase in mergers are likely overpowered by the remnant star formation \citep{rich+11}. IFS studies can delineate spatially between shock-dominated, and star formation dominated regions \citep{rich+15}, and might provide a path forward to identifying whether major mergers transit through a SPOG phase.

\begin{figure*}[t]
\centering
\includegraphics[width=\textwidth,clip,trim=1.5cm 0.2cm 2cm 0cm]{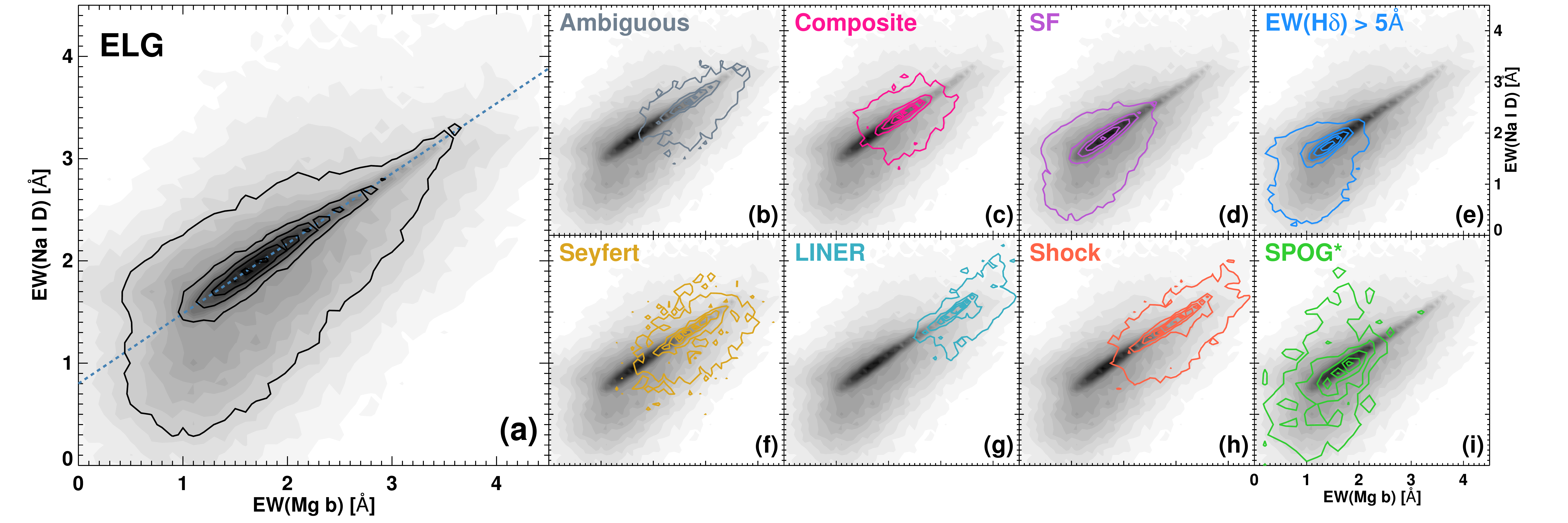}
\caption{\footnotesize \nad\ absorption is compared to Mg\,b for the ELG sample (a; black, underlying grayscale), with panels for ambiguous objects (b:\,gray), composites (c:\,pink), SF (d:\,purple), \ewhd\ objects (e:\,blue), Seyferts (f:\,yellow), LINERs (g:\,turquoise), objects within the shock boundaries (h:\,red), and SPOGs* (i:\,green). Contours for all subsamples are in increments of 10\% from the maximum in each distribution. In general, the Mg\,b and \nad\ absorption trace stellar population age \citep{vazdekis+10}, as is seen in the underlying parent sample. We see the general trend that LINERs contain the oldest stellar populations of the ELG sample, and \ewhd\ objects the youngest.  SPOGs* are the only population that has a non-negligible representation of galaxies with much larger \nad\ widths than Mg\,b, consistent with neutral winds \citep{rupke+05,murray+07,jeong+13}.}
\label{fig:NaD}
\end{figure*}

\begin{figure*}
\centering
\includegraphics[width=0.49\textwidth]{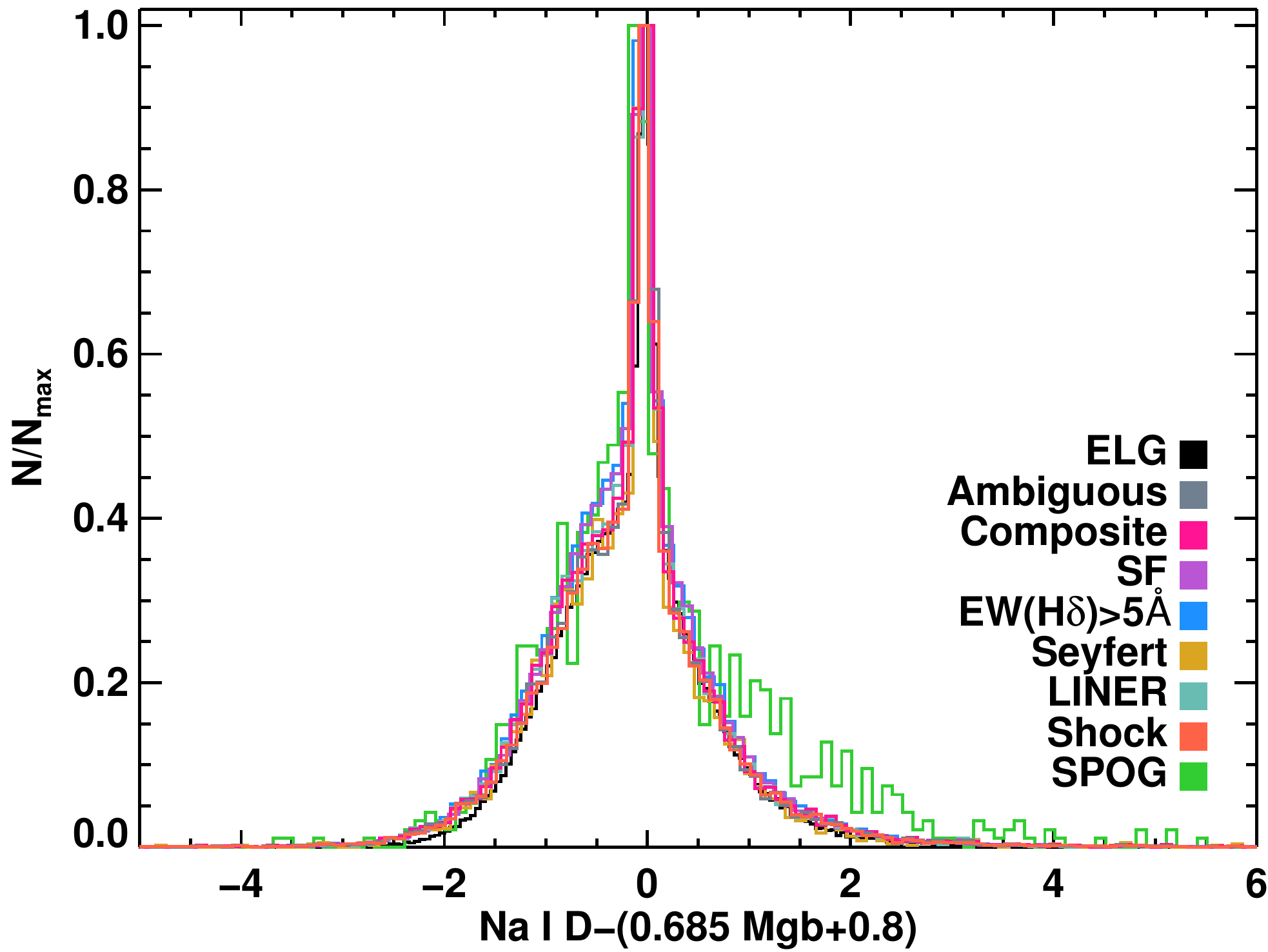}
\includegraphics[width=0.49\textwidth]{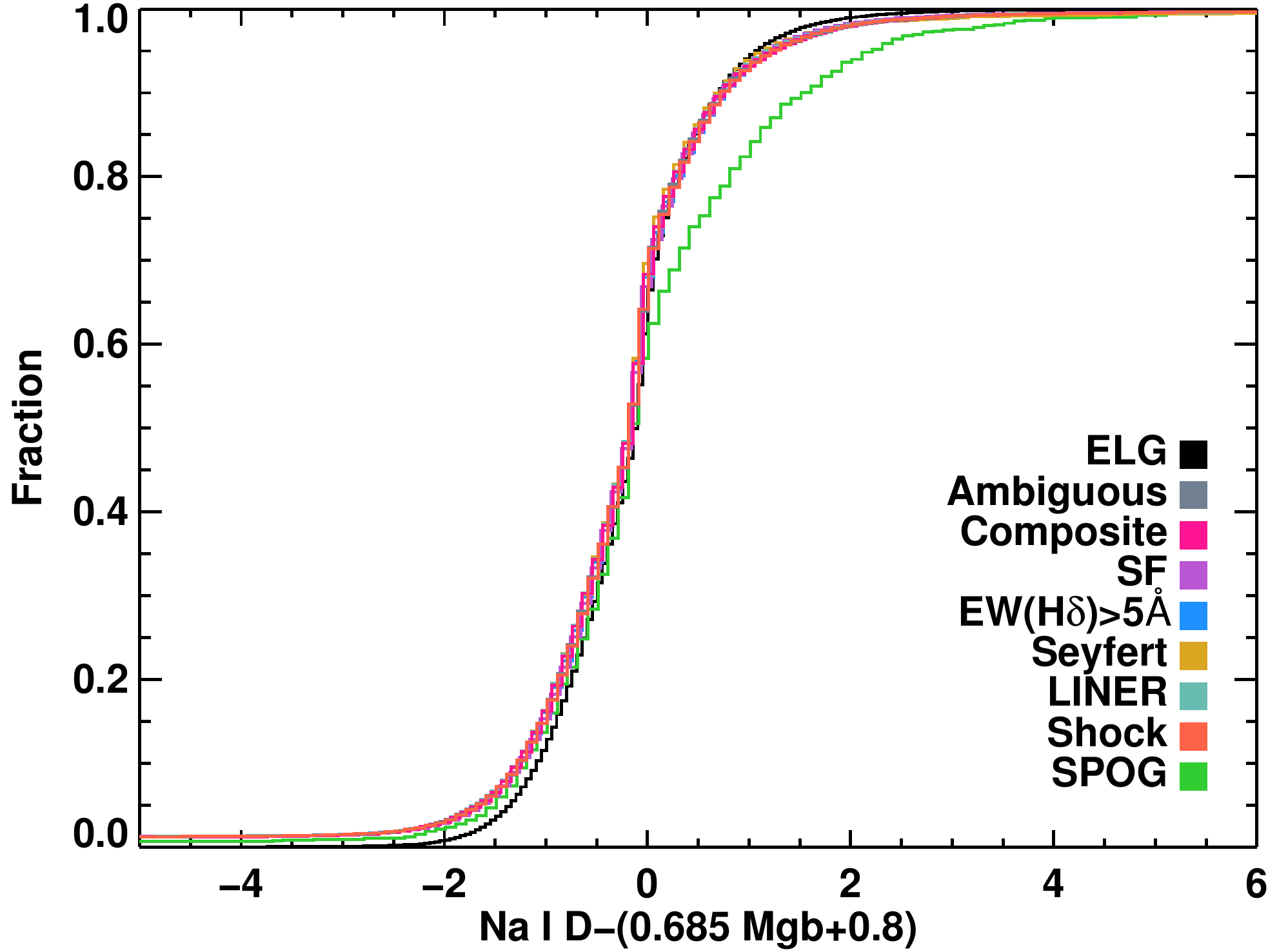}
\caption{\footnotesize The probability distribution function (left) and the cumulative distribution function (right) for the relative \nad\ versus Mg\,b in our parent sample (ELG) and subsamples, collapsed along the EW(Na\,D) = 0.685*EW(Mg\,b)+0.8 line shown in Figure\,\ref{fig:NaD}. SPOGs* show a clear departure from all other distributions, with a larger fraction of objects that are \nad-strong for their Mg\,b (23\% are beyond 1 standard deviation off of the ELG-derived relation), indicative of \nad\ winds \citep{murray+07}.}
\label{fig:NaD_cdf}
\end{figure*}

Shocks are also able to increase the molecular gas depletion time, and therefore the star formation quenching time, in galaxies by inhibiting star formation.  Shocks introduce excess kinetic energy that counteracts the gravitational instabilities within the gas that regulates star formation, leading to longer molecular gas depletion times. This phenomenon has been seen in AGN outflow hosts \citep{a15_sfsupp,aalto+16}, radio galaxies \citep{guillard+15,lanz+16}, and warm H$_2$-bright Hickson Compact Group galaxies \citep{guillard+12,a14_hcg57,a15_hcgco}. Shock-driven turbulence produced as a result of cyclical AGN episodes is one of the mechanisms capable of regulating star formation by rendering molecular gas infertile \citep{a14_stelpop}, possibly extending the lifetime of the SPOG phase. If shocks are able to regulate star formation in SPOGs, this suggests that the SPOG phase could include multiple short-duration SPOG episodes over a longer timescale (such as the A-star lifetime). A similar effect has been suggested for AGNs \citep{schawinski+15}.

Regardless of the demographics of SPOGs, in order to detect poststarburst spectral features, the star formation quenching timescale must be rapid. It is unclear whether a galaxy undergoing a SPOG phase is transforming via a quietly quenching mechanism (such as a minor merger or cosmic gas starvation), a major merger, cluster encroachment, group evolution, or all of these on $\lesssim$ 100\,Myr timescales.  An in-depth study of the environments and morphologies, including deep observations looking for stellar or H\,{\sc i} tidal streams would aid in constraining the fraction of each of these types of transformations that undergoes a detectable SPOG phase.





\subsection{Interstellar \nad\ in SPOGs*: evidence of winds?}
\label{sec:NaD}

The neutral sodium doublet (\hbox{\nad$\lambda$5890,\,5896}) is a stellar absorption feature in the atmospheres of cool stars \citep{bruzual+charlot03,vazdekis+10} and a well-known interstellar medium absorption line. In the past few decades, interstellar \nad\ has been detected in galactic winds in both starburst systems \citep{heckman+00,rupke+05,martin06,chen+10,sarzi+16} and AGN-driven outflow systems \citep{forster+95,krug+10,davis+12,rupke+15}.

Figure\,\ref{fig:NaD} shows the \nad\ versus Mg\,b absorption in ELG sample and subsamples.  There is a tight relationship between Mg\,b and \nad\ absorption throughout the ELG and most subsamples, consistent with the majority of the \nad\ absorption having a stellar origin. The \nad\ deficit at small Mg\,b that appears in the star-forming, \ewhd, and ELG samples is likely caused by in-filling of both the Na\,{\sc i} and nearby He\,{\sc i} emission lines, which is strongest in young and metal poor stellar populations.

An empirical relationship between Mg\,b and \nad\ in the ELG can be derived, resulting in the following equation: 

\begin{equation} \label{eqn:NaD}
{\rm EW(Na\,I\,D) = 0.685*EW(Mg\,b)+0.8}
\end{equation}

\noindent Most of the ELG subsamples trace the \nad--Mg\,b relation of Equation~\ref{eqn:NaD} (shown as a dashed blue line on Fig.~\ref{fig:NaD}a). SPOGs* are the marked exception.  The \nad--Mg\,b relation in SPOGs* most closely resembles the relation in \ewhd\ objects, but only at the small \nad\ end.  SPOGs* have a substantial (and unique) tail into strong \nad\ absorption (compared to Mg\,b).  Figure\,\ref{fig:NaD_cdf} shows the probability distribution function (PDF; left) and the cumulative distribution function (CDF; right) of the \nad--Mg\,b relation, collapsed along the line described by Equation~\ref{eqn:NaD}. While most subsamples follow the same distribution around Equation~\ref{eqn:NaD}, SPOGs* show a deviation, evident in both the PDF and the CDF. A non-parametric Kolmogorov-Smirnov test\footnote{\href{http://idlastro.gsfc.nasa.gov/ftp/pro/math/kstwo.pro}{http://idlastro.gsfc.nasa.gov/ftp/pro/math/kstwo.pro}} was run to compare SPOGs* to the other distributions, with the null hypothesis being ruled out with a confidence {\em p}\,$\ll$\,0.001. In detail, 245 (23\%) SPOGs* are at least 1 standard deviation (1$\sigma$) above Equation~\ref{eqn:NaD}, 120 (11\%) at least 2$\sigma$, and 56 (5\%) SPOGs* exceed Equation~\ref{eqn:NaD} by at least 3$\sigma$, confirming that a substantial population of SPOGs* show significant \nad\ enhancement above what is expected from stellar contributions based on Mg\,b absorption.

Enhanced \nad\ absorption is also observed due to a varying initial mass function (IMF) for the stars \citep{jeong+13,mcconnell+15}, but this has only been seen in the most massive, oldest systems. SPOGs* span a large range of masses, thus we do not believe that a varying IMF is the cause for the enhanced \nad. Another possibility is that there is likely a nonstellar contribution to the \nad\ line. The presence of interstellar \nad\ absorption amongst the SPOG* sample would suggest the presence of neutral winds, driving a non-negligible amount of interstellar medium (ISM) out of the system, thus aiding in the galaxy's transition.

Enhanced \nad\ absorption has also been shown to be a signature of neutral winds \citep{rupke+05}. It is possible that other subsamples may also host winds (such as the star-forming subsample; \citealt{sarzi+16}).  Given the enhancement in \nad\ absorption over Mg\,b, SPOGs* may have proportionally more interstellar neutral material as a population than the other subsamples.  Additionally, given the masses and A-star populations seen in SPOGs*, the enhanced \nad\ is much more likely to be due to interstellar winds.  \nad\ winds found in AGN-driven molecular outflows, such as Markarian\,231 \citep{rupke+11} and NGC\,1266 \citep{davis+12} were concentrated in the center, and traced the molecular outflow from the AGN. The velocity dynamics in both these systems made it clear that this absorption is due to the AGN-driven neutral wind.  Given that the discovery of NGC\,1266 was the catalyst for the SPOG survey, finding that SPOGs* as a population also have \nad\ absorption enhanced beyond what standard stellar population models predict is promising, and is a sign that our selection criteria have indeed produced the special population of transitioning galaxies that we were looking for.  

This shows that SPOGs* satisfying our \ewhd\ and shock criteria are not a random subsample of ELGs; they have distinct \nad\ behavior, showing a large wing with enhanced \nad\ absorption. In starbursts, these neutral winds tend to be attributed to stellar feedback, but given that the SPOGs* selection actively excludes star formation, the enhanced \nad\ absorption is likely from another source, such as AGN feedback. One potential caveat is that a \nad\ enhancement alone is not sufficient to confirm the presence of neutral winds, as enhanced \nad\ can also come from tidal debris sitting in front of the stars. Thus, an in-depth study of the kinematics and extent of the \nad\ absorption is necessary to determine the mass output and to confirm whether the \nad\ in SPOGs* is due to interstellar winds. A full investigation of the \nad\ properties of SPOGs* will be presented in a future paper.

\begin{figure*}[t]
\centering
\includegraphics[width=\textwidth,clip,trim=0.6cm 0cm 3cm 0cm]{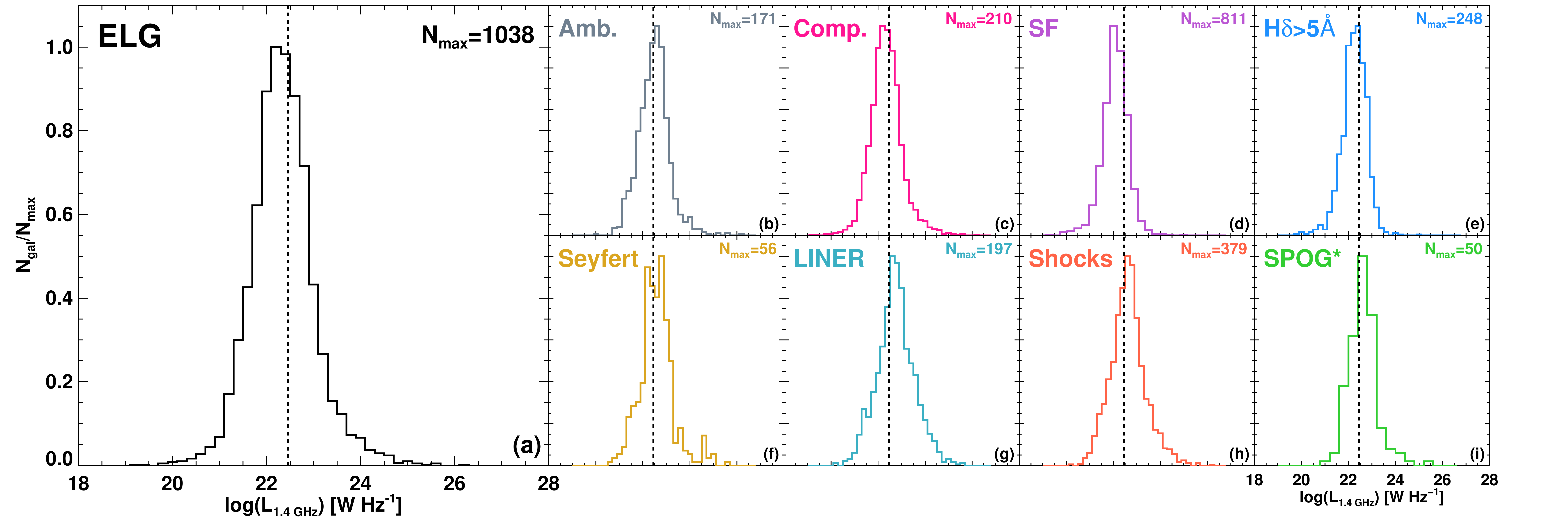}
\caption{\footnotesize Radio luminosity histogram of FIRST-detected objects in the ELG sample (a; black), ambiguous objects (b:\,gray), composites (c:\,pink), SF (d:\,purple), \ewhd\ objects (e:\,blue), Seyferts (f:\,yellow), LINERs (g:\,turquoise), objects within the shock boundaries (h:\,red), and SPOGs* (i:\,green). The dotted black line represents the mean radio luminosity of all ELG objects. N$_{\rm max}$ represents the number of objects in the most populated bin. The average $L_{\rm 1.4\,GHz}$ for SF and \ewhd\ objects is smaller than in LINERs, Seyferts, shocks, and SPOGs*.}
\label{fig:Lradio}
\end{figure*}

\begin{figure}[t]
\centering
\includegraphics[width=0.48\textwidth,clip,trim=0cm 0cm 0cm 0cm]{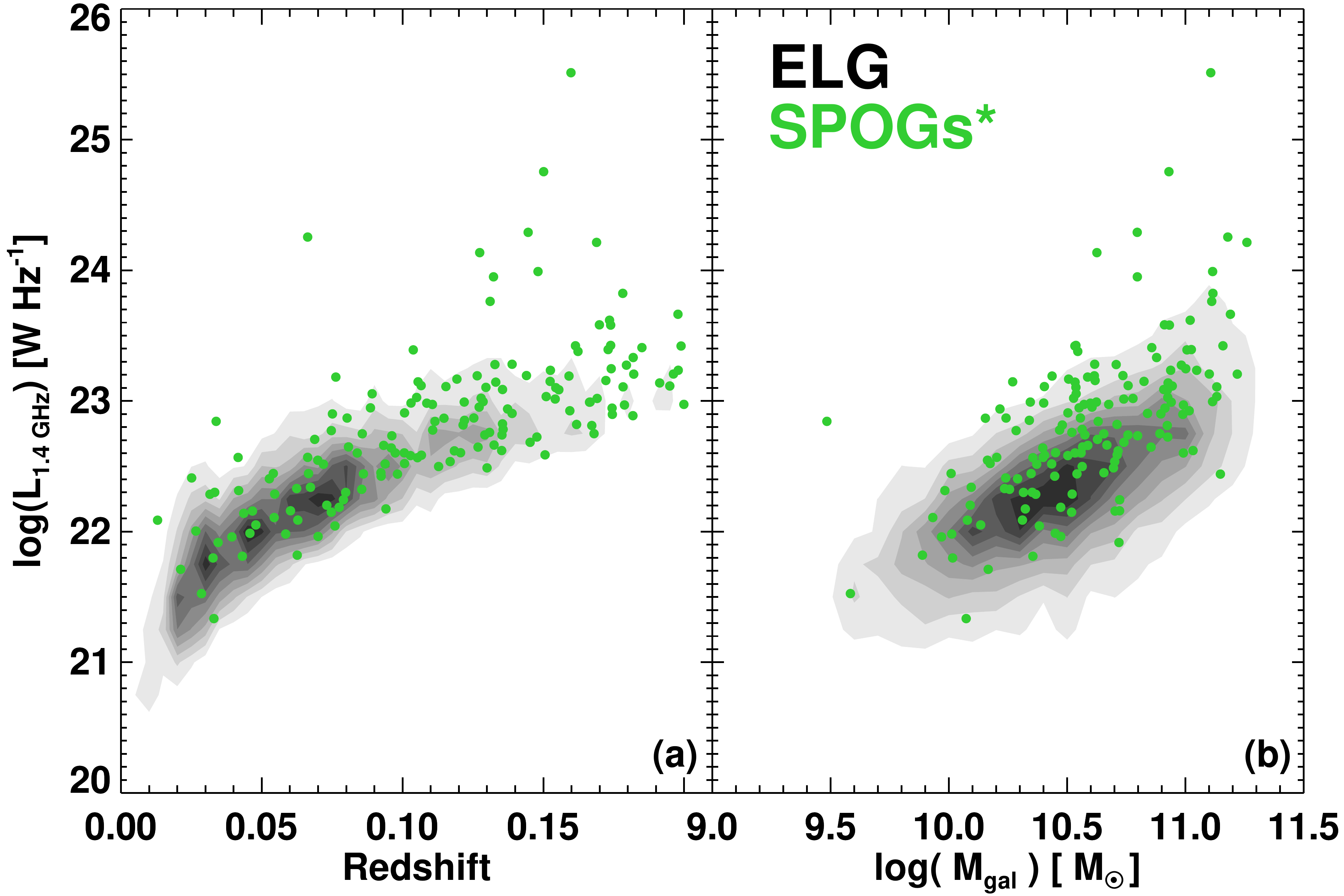}
\caption{\footnotesize Radio luminosity FIRST-detected objects in the ELG sample (black) overlaid with SPOGs* (green points) as a function of redshift (left; a) and mass (right; b). Contours levels are in increments of 10\% of the maximum.}
\label{fig:Lradio_spogs}
\end{figure}
\begin{table}[t]
\centering
\caption{FIRST detections} \vspace{1mm}
\begin{tabular}{l r r r}
\hline \hline
{\bf Type} & {\bf Number} & {\bf Percent}$^\dagger$ & $\mathbf{\langle \log(L_{\rm 1.4GHz}) \rangle}^\diamond$ \\
\hline
Ambiguous & 1,000 & 5.9 & 22.574$\pm$0.019\\
Composite & 1,127 & 7.9 & 22.364$\pm$0.014\\
SF & 2,362 & 2.1 & 22.212$\pm$0.010\\
EW(H$\delta$)\,$>$5\AA & 1,136 & 2.4 & 22.342$\pm$0.014\\
Seyfert & 349 & 7.3 & 22.687$\pm$0.037\\
LINER & 1,513 & 13.4 & 22.853$\pm$0.020\\
Shocks & 2,625 & 8.7 & 22.731$\pm$0.014\\
SPOGs* & 143 & 13.4 & 22.823$\pm$0.047\\
ELG & 6,351 & 4.0 & 22.450$\pm$0.008\\
\hline \hline
\end{tabular} \\
\label{tab:radio}
\raggedright {\footnotesize
$^\dagger$Percent of specific subsamples with FIRST matches\\
$^\diamond$The mean log 1.4\,GHz luminosities with standard deviation of the mean.
}
\vspace{1mm}
\end{table}

\subsection{Radio properties of SPOGs*}
\label{sec:radio}
Feedback via the injection of turbulent energy driven by a low-power radio AGN was a key aspect of the discovery of NGC\,1266 \citep{a15_sfsupp}. It is therefore of great interest to consider the potential influence of radio AGN feedback among SPOGs. We estimate the fraction of candidate ``radio-loud'' AGN hosts in our sample of SPOGs using the \citet{best+12} catalog of SDSS sources from DR7 with radio counterparts in the NRAO Very Large Array (VLA) Sky Survey \citep{nvss} and the Faint Images of the Radio Sky at Twenty centimeters (FIRST; \citealt{first}) survey. The \citet{best+12} catalog contains 18,286 radio-loud sources with flux densities measured at 1.4\,GHz that also have SDSS counterparts down to a limiting flux density of 5\,mJy. Of the 1067 SPOGs in our sample, only 32 ($\approx$\,3.0\%) are included in the \citet{best+12cat} catalog of SDSS radio sources.  To provide further insights into the population of faint radio sources with counterparts in our ELG catalog, we directly cross-matched this catalog with FIRST (with a flux limit of 1\,mJy) using {\sc topcat} \citep{topcat}.  We followed the strategy of \citet{ivezic+02}, who argued that a search radius of 1.5$^{\prime \prime}$ yields the best compromise among completeness and contamination (85\% and 3\%, respectively) when cross-matching SDSS galaxies with FIRST.  Using this search radius, the position of a radio source should lie within 3$^{\prime \prime}$ of the optical position of a galaxy from our ELG catalog. Of the 159,387 objects in the ELG sample, 6,351 (4\%) have FIRST matches. Table~\ref{tab:radio} provides information on the 1.4\,GHz detection rates for each subsample of objects within the ELG catalog. Figure~\ref{fig:Lradio} shows the distributions of 1.4\,GHz radio luminosities within the ELG sample and line diagnostic subsamples. LINERs, Seyferts, shocks, and SPOGs* have more significant luminosities than the star formation and \ewhd\ objects, given the shift seen in the radio luminosity distribution compared with the mean of the whole ELG sample (also shown in Table~\ref{tab:radio}). Figure~\ref{fig:Lradio_spogs} shows the archival 1.4\,GHz luminosity dependence on redshift and mass for the ELG sample.  Radio-detected objects span all redshifts and stellar masses.  A detailed statistical analysis of the radio continuum properties of the ELG sample and SPOGS in particular, that addresses the inherent distance and galaxy mass biases inherent to flux-limited surveys such as FIRST (e.g., \citealt{best+05}), will be presented in a future study.

The subsamples with the highest radio detection rates are SPOGs* and LINERs (both 13\%), significantly exceeding the Seyfert classified objects (7\%).  The high fraction of radio-detected LINERs is expected based on previous studies showing they are often hosted by massive galaxies with radio-loud AGNs \citep{owen+89}.  We speculate that the similarly high incidence of radio sources in SPOGs* could be a sign that these objects also host radio-loud AGNs, however, we emphasize that the detection of a weak radio source does not guarantee that an AGN present since star formation may also produce radio emission at low levels \citep{condon92}.  In addition to synchrotron emission from AGNs and recent star formation, shocks are known to produce radio continuum emission as well \citep{lisenfeld+10}. Thus, determining which mechanism dominates the energetic output of a galaxy is challenging.  Studies investigating the radio-22$\mu$m correlation (e.g., \citealt{appleton+04}) in the ELG sample to help identify objects in which AGN emission dominates at radio frequencies will be presented in future work.

\subsection{Potential catalysts of the SPOG phase}
Radio jets could explain many observational properties of SPOGs: their short timescales, their lack of ongoing star formation, and the fact that they harbor powerful sources of ionization. Jets from radio galaxies can inject significant turbulence (traceable via shocks) into a galaxy's ISM, thereby inhibiting its ability to form stars \citep{lanz+16}. Strong radio galaxies have been found to be cyclical through the discovery of fossil radio shells \citep{schoenmakers+00}, extending the radio bright timescale beyond the radio jet timescale ($\sim$\,10\,Myr; \citealt{turner+15}). The quintessential SPOG NGC\,1266 is a case study for this phenomenon, with turbulent energy injection from the radio jets powering the shocks \citep{alatalo+11,a14_stelpop}. The fact that SPOGs* show slightly elevated radio detections compared to the rest of the ELG (and its subsamples; see \S\ref{sec:radio}) seems to indicate the possibility that radio jets could play a role in transitioning galaxies, but confirming this requires deeper and higher resolution radio data.

Mergers are thought to play a role in the morphological and color transition of poststarburst galaxies and these interactions could also fuel black hole growth in the form of quasar activity lasting up to $<$\,10$^8$\,yr \citep{canalizo+00,martini04,mullaney+12,canalizo+13}. In order to be as massive as they are, all supermassive black holes are expected to go through powerful phases of accretion. Several studies have made connections between AGN/quasar phases and poststarburst activity \citep{wild+09,brown+09,kocevski+11,cales+11,cales+13,yesuf+14,cales+15}, and though circumstantial, SPOGs could be the smoking gun revealing the AGN-host galaxy connection. Due to our shock boundaries, we are not sensitive to AGN activity producing the hardest radiation fields (\oiii/H$\beta$\,$>$\,1.03), though we would be able to identify objects with softer radiation fields (as is seen in some Type\,II quasars; \citealt{villar-martin+08}).\footnote{A secondary check that can be run on this population is searching for substantial \oiii\ emission, which should separate the quasars from shock-dominated objects.}  The shock boundary might also allow us to observe systems that have recently undergone quasar activity, given both the stellar populations of quasars found by \citet{canalizo+13}, as AGN ``flickering'' time variability arguments of \citet{schawinski+15} suggest that the short timescale AGN events could be caught using the SPOG criteria.

Galaxies falling into clusters or in group environments are also known to be transforming, having rapidly truncated their star formation \citep{dressler80,dressler+gunn83,johnson+07,ko+13,lee+15}. Many of these systems have also been detected to have shocks traced by warm H$_2$ \citep{sivanandam+10,cluver+13}. Given the simultaneity of shocked gas and truncated star formation, shocked group galaxies and cluster galaxies are likely to be identified using the SPOG criterion, and some examples are very likely present in our sample.

It is possible that a subset of SPOGs are not transitioning from blue to red at all, but instead are being refueled by the accretion of cold gas, temporarily replenishing star formation \citep{kannappan+09,mcintosh+14} and producing shocks due to the interaction between the accreted material and the in-situ ISM of the recipient galaxy. This catalyst was argued by \citet{dressler+13} as the origin of poststarburst galaxies in both clusters as well as the field, having transitioned across the green valley in the opposite sense (though this phase is expected to be short-lived; \citealt{young+14,appleton+14}).

Most likely, the SPOG criterion has identified examples of all of these sub-populations of transitioning galaxies: some hosting AGN-driven outflows, some on the tail-end of merging, some transitioning in a group or cluster setting, some replenishing from accretion, and likely some that do not fit into any of these categories. It will take detailed studies of their environments, morphologies, interaction histories, kinematics, interstellar media, star formation histories, and AGN properties to understand the nature of objects in a SPOG phase.

In order to gain deeper understanding of SPOGs, we then must compare how SPOGs relate to the broader class of quenching galaxies which may or may not have shocks. SPOGs* as a sample is an excellent starting point, pinpointing the rare objects that warrant further study, and opening the door to a deeper understanding of the initial conditions that trigger the transformation of galaxies from blue spirals to red early-types (and possibly back again). \\

\section{Summary}
\label{sec:sum}

As galaxies age, they move from the blue cloud (star-forming spirals) to the red sequence (quiescent early-types) in color while also transforming morphologically.  A few case studies in the nearby universe seem to indicate that a population of galaxies are transitioning ``quietly'', with few outward signs of the transformation, and warrant further investigation. Using the OSSY catalog from SDSS DR7 and requiring robust detections of all diagnostic emission lines (the ELG sample), we combine Balmer absorption selection criteria with emission line ratios consistent with shocked emission (but excluding pure star formation) to create the Shocked POststarburst Galaxy Survey (SPOGS). However, these objects can only be considered candidates until observations confirming the presence of shocks can take place. We summarize our findings below. 

\begin{itemize} 
\item Traditional poststarburst searches, which use the presence of Balmer absorption and the absence of nebular ([O\,{\sc ii}] and/or H$\alpha$) emission, are able to identify galaxies that have recently transformed morphologically, but miss a large subset of transitioning galaxies in which other mechanisms (such as shocks and AGNs) excite the [O\,{\sc ii}] and H$\alpha$.

\item We show that 0.2\% (1,067 galaxies) of the OSSY continuum sample fit the SPOGs criteria, which is comparable to catalogs of completely passive poststarburst galaxies, indicating that galaxies harboring powerful sources of ionization from AGN, LINERs and/or shocks are also an important contributor to the poststarburst class. SPOGs* are seen to be in the green valley, though with bluer colors than E+A-selected OSSY galaxies, suggestive of being in an earlier stage of transition.


\item The \nad\ properties of SPOGs* are unique amongst the subsamples studied, showing a significant population of enhanced \nad\ objects, suggesting that SPOGs* contain interstellar \nad, which might imply the presence of galactic winds.

\item The SPOG* subsample has a 13\% radio detection rate (the highest in the ELG, along with LINERs), suggesting that many SPOGs* host AGNs, although other origins for the radio emission must be ruled out first.

\item It is likely that the SPOG criteria have sampled a heterogeneous set of transitioning objects, including those with AGN-driven outflows, objects undergoing mergers, quasar hosts, galaxies entering clusters, in group environments, replenished early-type galaxies, as well as none of the above.  Further studies of this SPOG* sample will be able to confirm whether the emission line ratios trace shocks, sample the morphologies and environments that SPOGs* inhabit, measure the AGN power and star formation rate, and determine the interstellar medium properties of objects undergoing a SPOG phase, to put a limit on the relative importance of each catalyst to the SPOG phenomenon.
\end{itemize}


\acknowledgments KA and SLC thank the thorough, thoughtful and expert recommendations from the anonymous referee, which have vastly improved this manuscript. KA also thanks Alan Dressler for conversations about previously made assumptions, adding further depth to the manuscript. KA is supported through Hubble Fellowship grant \hbox{\#HST-HF2-51352.001} awarded by the Space Telescope Science Institute, which is operated by the Association of Universities for Research in Astronomy, Inc., for NASA, under contract NAS5-26555. SLC was supported by ALMA-CONICYT program 31110020.  Partial support was provided  to KA and PNA by NASA observations through a contract issued by the Jet Propulsion Laboratory, California Institute of Technology under a contract with NASA. KN acknowledges support from NASA through the {\em Spitzer} Space Telescope. AMM and LJK acknowledge the support of the Australian Research Council (ARC) through Discovery project DP130103925.

Funding for SDSS-III has been provided by the Alfred P. Sloan Foundation, the Participating Institutions, the National Science Foundation, and the U.S. Department of Energy Office of Science. The SDSS-III web site is http://www.sdss3.org/.  The National Radio Astronomy Observatory is a facility of the National Science Foundation operated under cooperative agreement by Associated Universities, Inc.  This research has made use of the NASA/IPAC Extragalactic Database (NED) which is operated by the Jet Propulsion Laboratory, California Institute of Technology, under contract with the National Aeronautics and Space Administration. 

\bibliographystyle{aasjournal}
\bibliography{../../master}

\begin{appendix}
\section{Investigating the potential biases for the ELG and SPOGs}
\label{app:bias}
\begin{figure*}[t]
\centering
\subfigure{\includegraphics[height=2.1in]{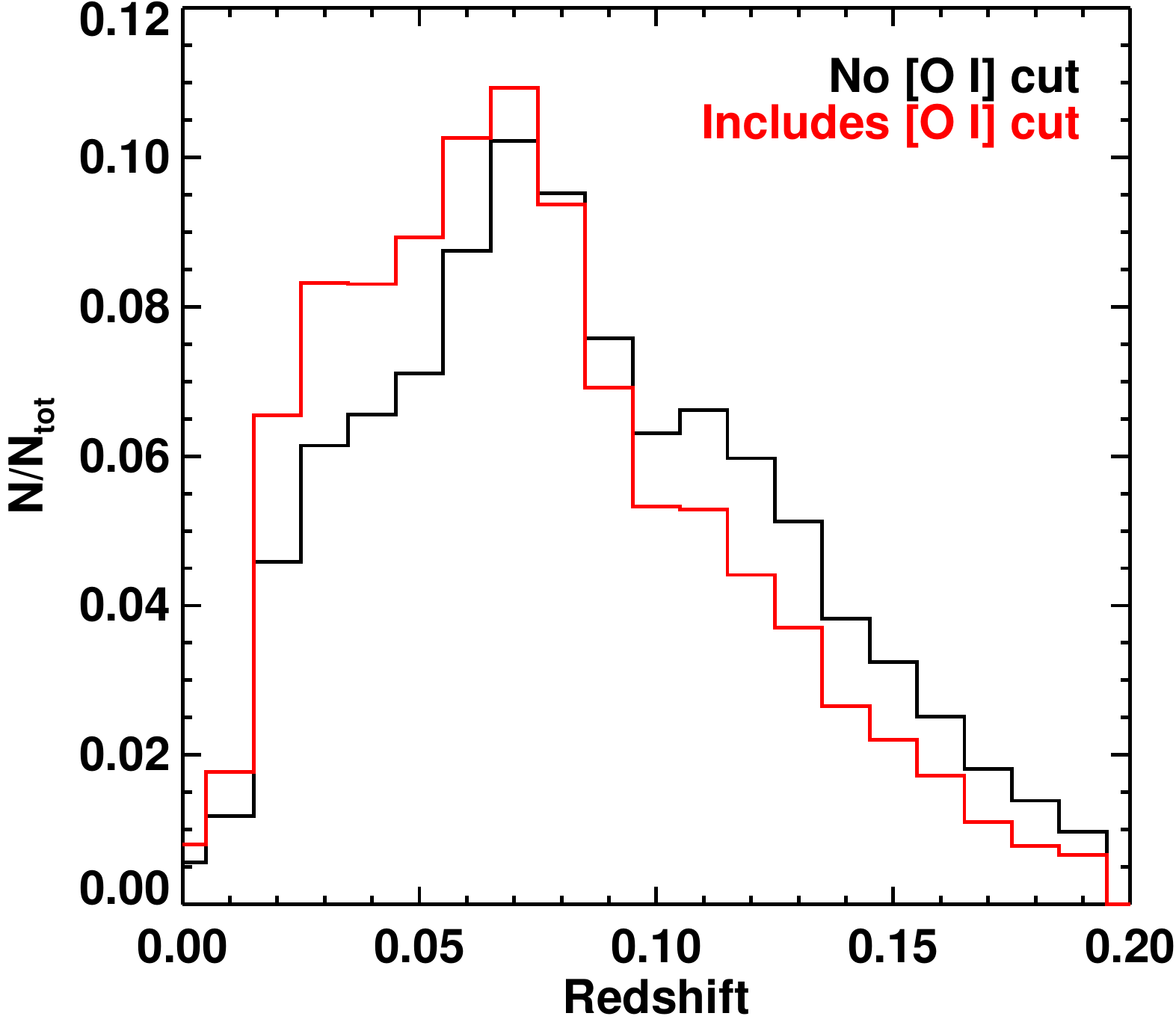}}
\subfigure{\includegraphics[height=2.1in]{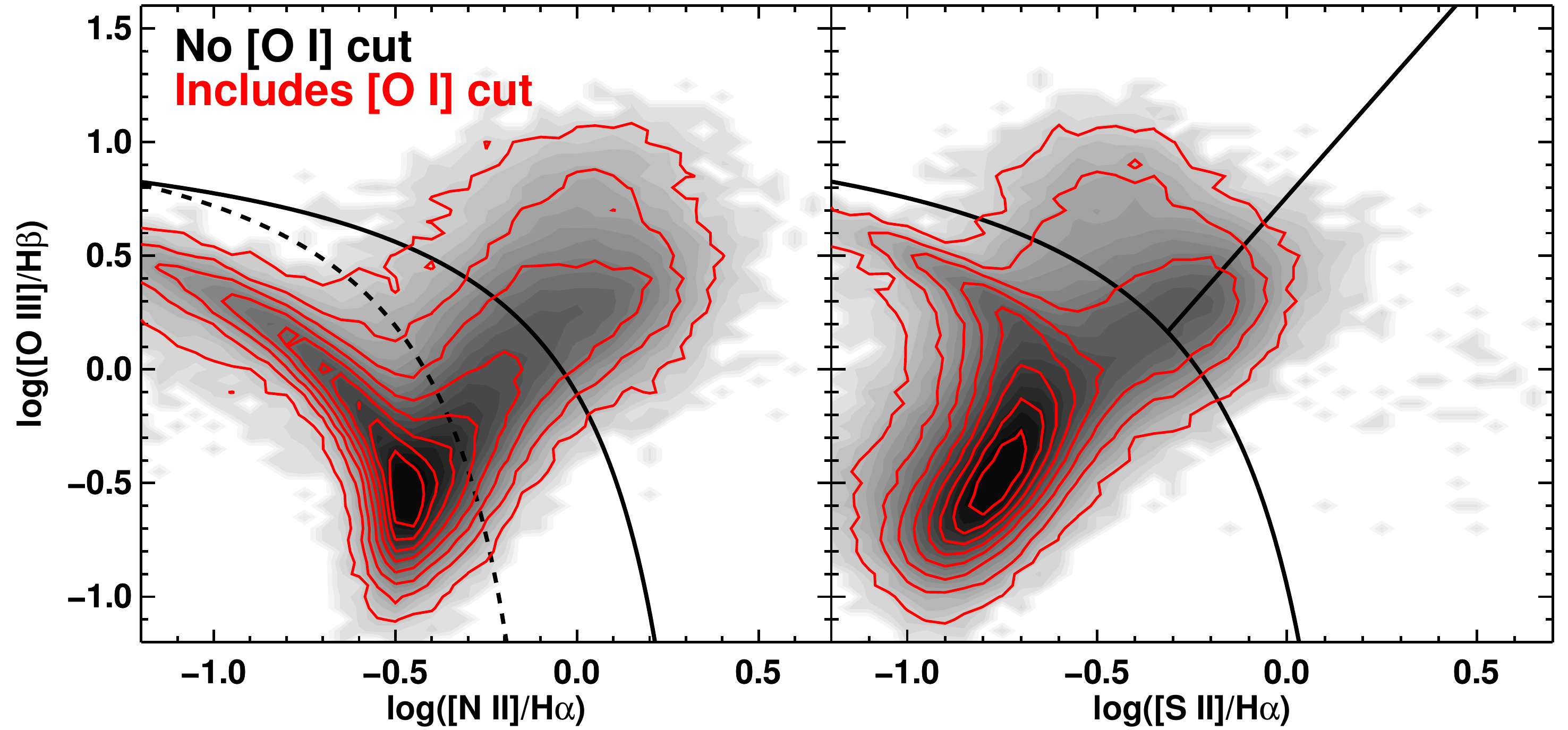}}
\caption{\footnotesize {\bf(Left):} The redshift distributions of OSSY galaxies that pass the continuum S/N cuts, comparing the galaxies detected in H$\alpha$, H$\beta$, \oiii, \sii, and \nii\ (black) to those that pass the \oi\ cut (red). {\bf(Right):} Comparison of the \nii/H$\alpha$ and \sii/H$\alpha$ for objects with (red) and without (grayscale) \oi\ cuts.}
\label{fig:oi_comp}
\end{figure*}
\begin{figure*}[t!]
\centering
\subfigure{\includegraphics[width=0.48\textwidth]{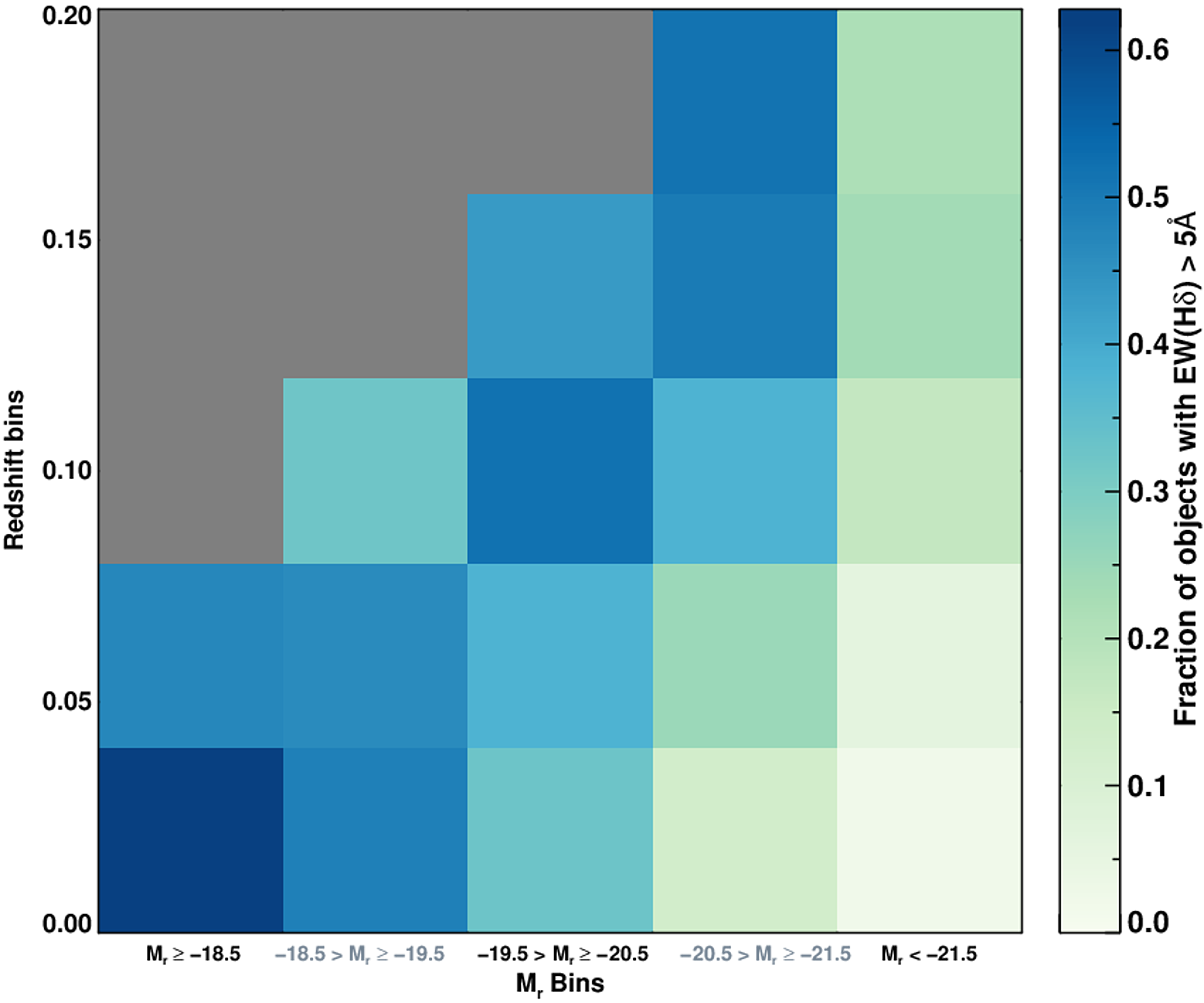}}
\subfigure{\includegraphics[width=0.48\textwidth]{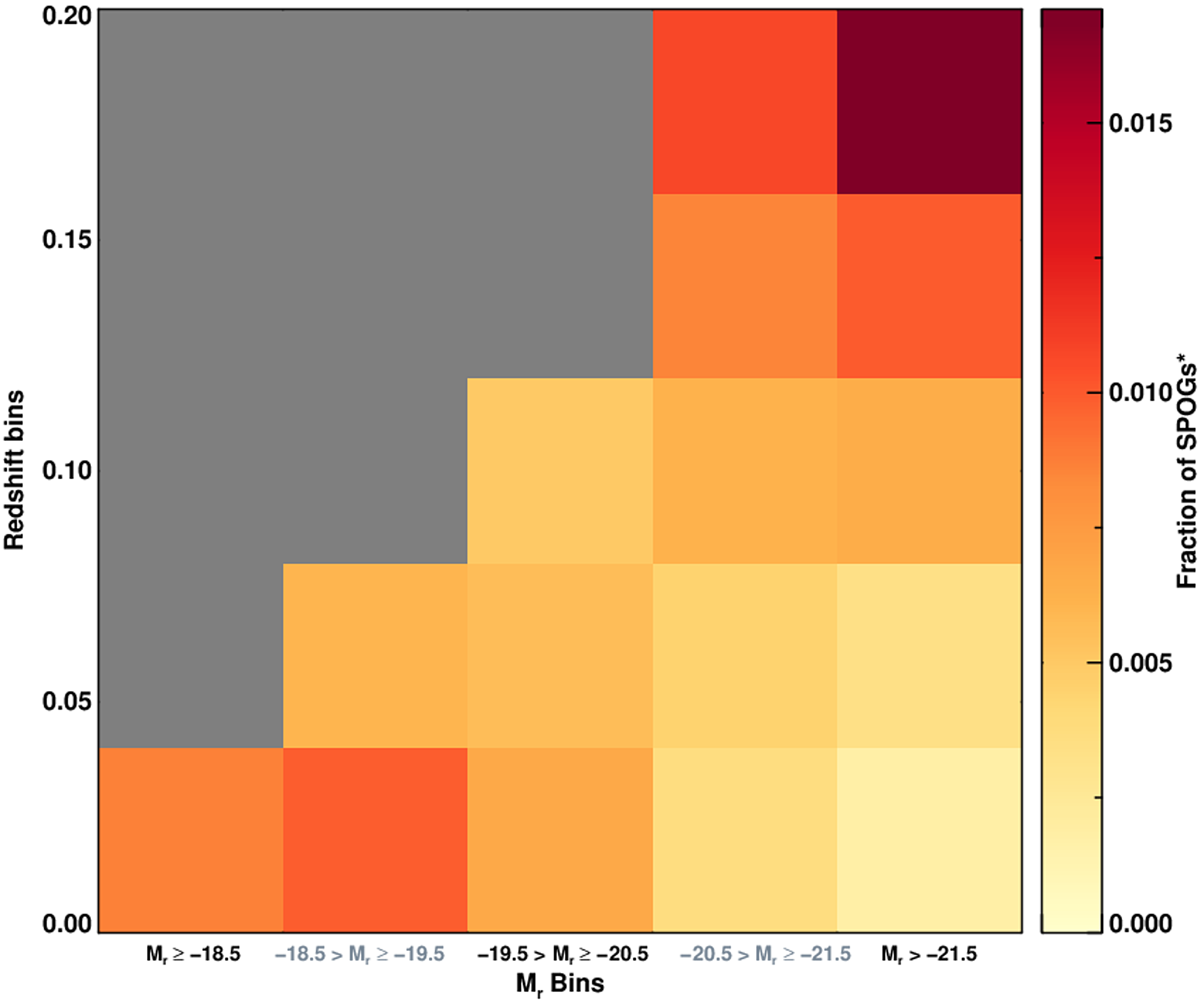}}
\caption{\footnotesize{\bf(Left):} Detections of galaxies with \ewhd\ as a function of absolute {\em r}-band magnitude and redshift. The fraction shows both a trend in $M_r$, and a trend in redshift, with the smallest $M_r$ lowest redshift bin containing the most objects with \ewhd, and the lowest detection rate in the largest $M_r$ bins. Regions shaded gray contain no objects. {\bf(Right):} The SPOG* fraction as a function of absolute {\em r}-band magnitude and redshift. Interestingly, the SPOG* detection rate does not appear to show trends with redshift below $M_r < -20.5$, though the overall detection rate does not vary by more than a factor of 3 per bin. The largest $M_r$, highest redshift bins have the highest SPOG* identification rate, with a secondary peak at low redshift, small $M_r$ objects.}
\label{fig:mass_z}
\end{figure*}

With the ability to study millions of galaxies over cosmic time, astronomers using SDSS data have transformed the study of galaxy evolution. However, despite this success, fiber spectroscopy suffers from a major drawback: for nearby galaxies, a fiber is only able to target its center.  Within the OSSY sample, the size of the 3$''$ SDSS fiber varies in physical scale upon the galaxy by over two orders of magnitude, based on the redshift that is being investigated. For this reason, we must discuss possible aperture effects that are present in both the ELG and the SPOG* sample. 

\subsection{The effect of requiring bright \oi\ emission to the sample}
\label{app:oi}
First, we explored whether the emission line selection criterion (and thus the relative fractions of diagnostic classifications) was biased by redshift, due to requiring the detection of a weak line (\oi). There were 280,378 objects that met the cuts in all other lines (H$\alpha$, H$\beta$, \nii, \sii, and \oiii). 

We plotted the distributions of redshifts in Fig.~\ref{fig:oi_comp}a, which shows that the \oi\ cut slightly favors lower redshift objects, consistent with a slight Malmquist bias (see: \S\ref{app:malmquist}). The average redshift of the objects where no \oi\ cut was applied is {\em z}\,=\,0.090, compared to \oi\ cuts, with {\em z}\,=\,0.080. 

Figure~\ref{fig:oi_comp}b shows the \nii/H$\alpha$ and \sii/H$\alpha$ vs \oiii/H$\beta$ line diagnostics without the \oi\ cut in grayscale, compared to applying the \oi\ cut (red). The \oi\ cut does not prefer specific segments of the line diagnostic diagrams. We therefore do not believe that our selection for the \oi\ line imposes biases significant enough to impact our conclusions.

\begin{figure*}[t!]
\centering
\includegraphics[width=\textwidth,clip,trim=1.5cm 0cm 0cm 0cm]{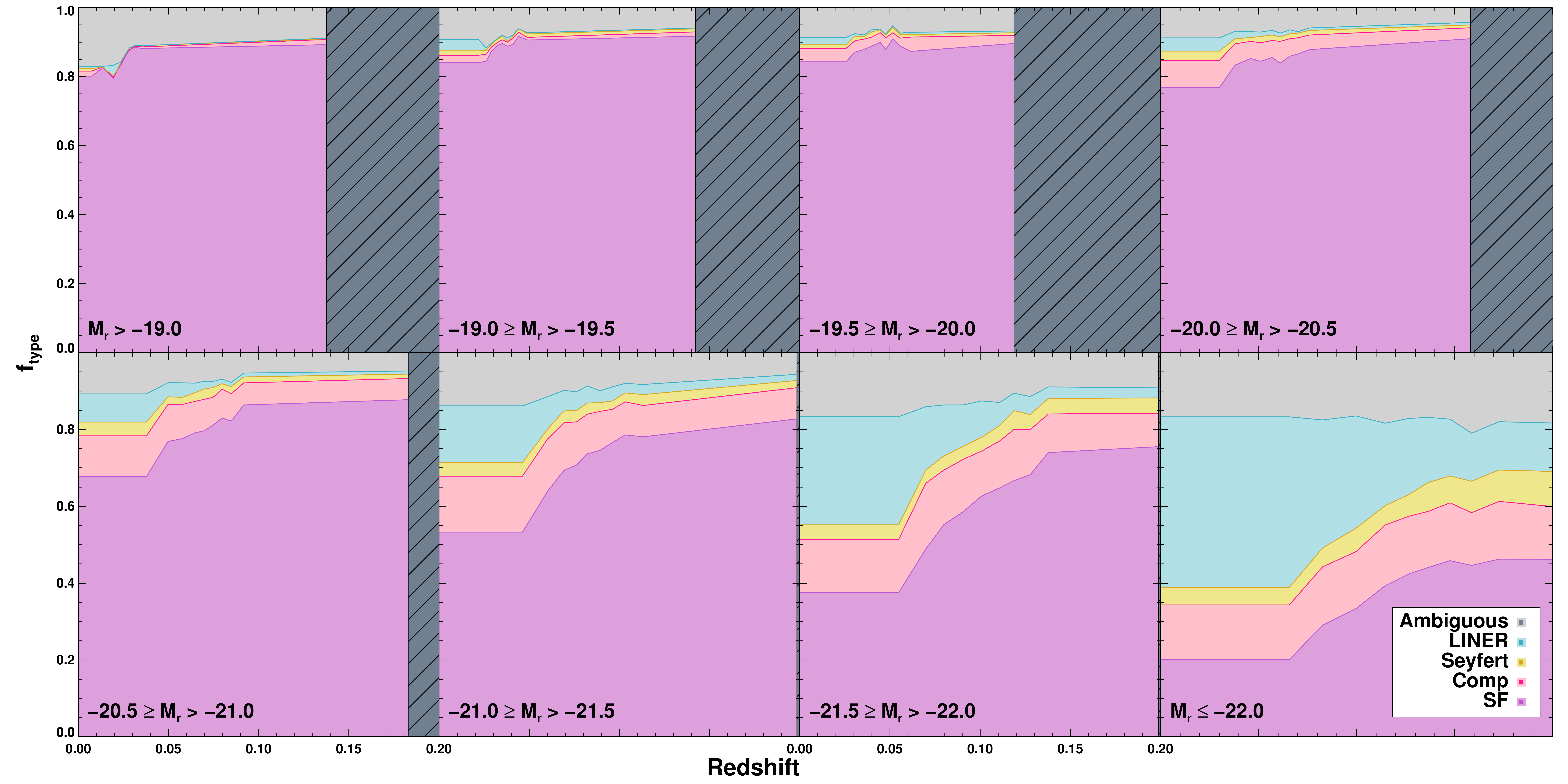}
\caption{\footnotesize The line diagnostic classifications from \citet{kewley+06} as a function of redshift in 8 distinct mass bins, from lowest mass to highest. The redshift bins were determined by sub-dividing all galaxies in each mass category evenly between 10 separate redshift bins. The redshift bin point is placed at the high reshift point of each bin. This means that all redshift bins have the same number of total objects. The dark gray blocked region of the plot represents redshift space where no ELG objects within the mass bin are detected. In the cases of the all but the highest mass bin, the majority of objects are star-forming, with a  significant minority falling into the ambiguous category.}
\label{fig:bpt_mass_z}
\end{figure*}

\subsection{Aperture and Malmquist Bias}
\label{app:malmquist}

The small central aperture did find objects classified as SPOGs* that had extended star formation (see \S\ref{sec:sf_contaminants}). To investigate the magnitude of this issue, we plotted the \ewhd\ and SPOG* detection rate as a function of $M_r$ and redshift, shown in Figure \ref{fig:mass_z} (note: Figure~\ref{fig:Mr_z} shows that small $M_r$, high redshift objects are not found in the ELG). The identification fraction of \ewhd\ objects strongly depends both on $M_r$ and redshift, with large $M_r$ objects being less likely to have strong Balmer absorption. At the smallest $M_r$, the \ewhd\ fraction decreases slightly with redshift, possibly due to a number of low mass dwarfs that fit the ELG criterion, but this trend reverses as $M_r$ grows larger, with increasing detection rates amongst the higher redshift ELG objects. It is very likely that this is due to the SDSS fiber picking up larger areas of the star-forming disk (as opposed to the bulge).

The SPOG* detection vs. $M_r$ and redshift (Fig.~\ref{fig:mass_z}b) has two peaks, one toward the larger $M_r$, highest redshift objects, and one near the smallest $M_r$, lowest redshift, though the SPOG* identification rate does not change by more than a factor of 3 through the sample.  While aperture effects could explain this peak, other effects (including Malmquist bias and metallicity effects) might also explain the slightly elevated detection rate in the small $M_r$, low redshift bin. 

Malmquist bias (the proclivity to sample the brightest objects; \citealt{malmquist25}) is well known in sample selections, especially those with flux cutoffs. The most obvious manifestation is seen in Figure~\ref{fig:Mr_z}, showing that the average absolute magnitude within the ELG increases with redshift. It is possible that the high detection rate of SPOGs* in the highest mass, highest redshift objects is in part due to Malmquist bias, with the most massive objects at the highest redshift also tending to sample a rarer, brighter subsample. In particular, the fact that there is a significant drop in the fraction of \ewhd\ objects at the high mass, high redshift end seems to indicate that the ELG is sampling other bright sources of emission, given that the high mass end of the galaxy distribution function has significantly fewer star-forming galaxies \citep{schawinski+14,ogle+16}. Given that Malmquist bias has manifested in the detection fractions of Balmer absorbing systems (and possible SPOGs), it may also impact the relative proportion of line diagnostic types that are detected within the ELG.

To investigate this, we plotted the redshift-dependent line diagnostic classification in Figure~\ref{fig:bpt_mass_z}. In most cases, the mass has a significant effect on the fractional representation of each line diagnostic classification. The 4 lowest mass bins contain at least 75\% star-forming classified objects, with a substantial fraction of objects also classified as ambiguous at the lowest redshifts. At these low redshifts, the ELG would be able to detect dwarf star-forming galaxies. These galaxies have low metallicities, which can pump the \oiii\ compared to H$\beta$ emission, and suppress \nii\ compared to H$\alpha$ emission, pushing the emission line diagnostic into the Seyfert class \citep{kewley+13b,kewley+13a}. This would then create a disagreement between the various line diagnostics, thus making the resultant integrated classification ambiguous.

AGN line classification appears to be consistent to {\em z}\,$<$\,0.3 \citep{lamassa+12}. The AGN fraction within the ELG does not change much with redshift over the different mass bins, although there are more AGNs identified as a function of absolute magnitude, consistent with observations that AGNs lie in massive host galaxies \citep{kauffmann+03}. The detection rate of composite objects also does not appear to change with redshift (and thus galactic area subtended by the SDSS fiber), consistent with the results of \citet{lamassa+13} suggesting that the circumnuclear star formation in AGN hosts tends to be compact ($<$1.7\,kpc).

In the two highest mass bins, there are some trends with redshift, including a larger percentage of objects with LINER classifications at low redshift. This is in agreement with the behavior of LINERs from \citet{kewley+06}, as LINER-like emission tends to be weak \citep{yan+06}, and located in massive galaxies \citep{sarzi+06,sarzi+10}, thus the more nearby massive LINER host galaxies are the most likely to surpass the emission line thresholds of the ELG sample. Since LINER emission is thought to be widespread in massive galaxies \citep{sarzi+06,sarzi+10,capetti+baldi11}, one might expect the detection rate to increase with redshift (as is seen in star-forming objects) due to the change in the physical scale of the SDSS fiber, but the intrinsic weakness of the emission seems to be a stronger effect in our LINER detection rate.

Overall, this likely means that the SPOG* catalog has notable biases, namely, a slight aperture bias and Malmquist bias. The aperture bias allows some low redshift objects to be classified as SPOGs* despite hosting extended star formation, though might also work in the other direction (as is seen with \ewhd\ objects in Figure~\ref{fig:mass_z}) when the SDSS fiber is able to sample larger portions of the star-forming disk (rather than purely the bulge light). The SPOG* sample also suffers from Malmquist bias, with the detection rate increasing as a function of both absolute magnitude and redshift, so SPOGs* have been more commonly found amongst the brightest and rarest of the objects in our sample. Despite these biases, the SPOG criterion seems to have uncovered a new sample of transitioning galaxies, though future surveys and applications of the SPOG catalog should keep these biases in mind.

\end{appendix}

\end{document}